\documentclass[journal=jacsat,manuscript=article]{achemso}
\usepackage[version=3]{mhchem} 
\usepackage{booktabs}   
\usepackage{siunitx}    
\usepackage{multirow}
\usepackage[T1]{fontenc}       
\usepackage[utf8]{inputenc}    
\usepackage{subfigure}

\author{Owais Ahmad}
\affiliation{Department of Materials Science and Engineering, Indian Institute of Technology Kanpur, Kanpur 208016, India}

\author{Aravind K}
\affiliation{Department of Materials Science and Engineering, Indian Institute of Technology Kanpur, Kanpur 208016, India}

\author{Naveen Kumar}
\affiliation{Department of Materials Science and Engineering, Indian Institute of Technology Kanpur, Kanpur 208016, India}
\alsoaffiliation{Department of Materials Engineering, Indian Institute of Science, Bangalore, Bengaluru 560012, India}

\author{T.A. Abinandanan}
\affiliation{Department of Materials Engineering, Indian Institute of Science, Bangalore, Bengaluru 560012, India}

\author{Somnath Bhowmick}
\email{bsomnath@iitk.ac.in}
\affiliation{Department of Materials Science and Engineering, Indian Institute of Technology Kanpur, Kanpur 208016, India}

\author{Rajdip Mukherjee}
\email{rajdipm@iitk.ac.in}
\affiliation{Department of Materials Science and Engineering, Indian Institute of Technology Kanpur, Kanpur 208016, India}

\title{Bridging Phase-Field Model and Deep Learning for Predicting 2D and 3D Microstructure Evolution in Ternary Alloys}

\begin{document}
\begin{tocentry}
\includegraphics[width=8.25cm, height=4cm]{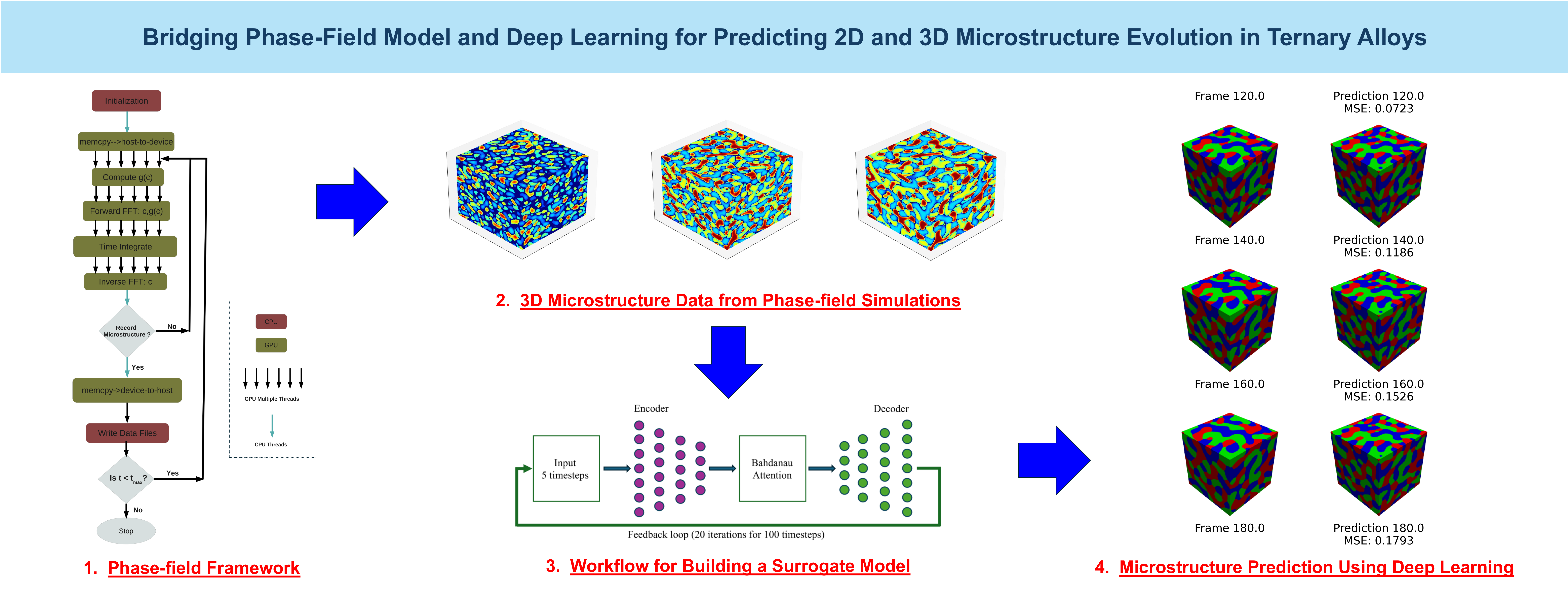}
\end{tocentry}

\begin{abstract}
We develop a hybrid framework that integrates a phase-field model (PFM) with an attention-enhanced deep learning (DL) architecture to study ternary spinodal dealloying, a sophisticated self-organization approach used to fabricate three-dimensional bicontinuous, hierarchical nanoporous materials for nanotemplate applications. The study captures three distinct phase-separation mechanisms that emerge during the early stages of spinodal decomposition in both two and three dimensions. The DL workflow consists of three key components: (i) a dimensionality-reducing autoencoder that provides compact representations of high-resolution microstructure images ($256{\times}256{\times}3$), (ii) an attention-augmented convolutional long short-term memory (ConvLSTM) network that learns complex spatiotemporal correlations governing microstructure evolution, and (iii) a novel slice-by-slice strategy that enables extension of the model to three-dimensional systems ($128{\times}128{\times}128{\times}3$). We further demonstrate a hybrid simulation strategy in which PFM accurately captures rapid early-stage microstructure evolution, while the DL model efficiently predicts late-stage coarsening dynamics. The trained DL model achieves remarkable predictive accuracy, maintaining fidelity up to 400 timesteps ahead and generalizing to compositions outside the training distribution. By bridging the physical fidelity of PFM with the computational efficiency of DL, this framework establishes a robust platform for predictive modeling of microstructure evolution in complex multicomponent systems.
\end{abstract}
\maketitle
\def\thefootnote{$\dagger$}\footnotetext{These authors contributed equally to this work}\def\thefootnote{\arabic{footnote}}
\section{Introduction}
\label{intro}

Spinodal decomposition (SD) is a process that causes a homogeneous mixture to separate into multiple phases and it has important practical applications in several systems, starting from metallic alloys to polymers~\cite{doi:10.1021/acsami.5c13603, doi:10.1021/acsami.4c03011}. 
SD in an A-B-C ternary and higher-order systems can lead to the formation of three-phase microstructures. More importantly, the three phases may be co-continuous, forming a tri-continuous microstructure. This significantly broadens the composition (or volume fraction) window in which a phase remains continuous and percolating, making higher-order alloy systems fundamentally different from their binary counterparts. Ternary three-phase spinodal decomposition plays a crucial role in generating complex, bicontinuous microstructures with tortuous solid-vapor interface during dealloying, where compositional fluctuations at interfaces lead to phase separation into interconnected domains \cite{Geslin2015}. In such systems, the interplay between interfacial spinodal decomposition and diffusion-coupled growth produces hierarchical architectures with tunable length scales, ranging from nanometers to micrometers \cite{Geslin2015}. Dealloying of ternary alloys, such as Al--Cu--Sn and Mg--Cu--Sn, proceeds as a selective corrosion process in which one or more less noble elements are dissolved, enabling the formation of nanoporous structures with controlled composition and morphology \cite{Zhang2018, Song2015}. This process may also involve simultaneous dealloying and realloying phenomena, which influence phase evolution and the stability of intermediate intermetallic phases~\cite{Song2015}. The resulting multiphase bicontinuous ligaments with nanoporous networks exhibit high surface area and efficient transport pathways, making them particularly attractive for catalytic and nanotemplate applications \cite{Qi2013}. Consequently, controlling ternary spinodal decomposition provides a pathway to engineer 2D and 3D microstructures with tailored topology and functionality for advanced nanotemplate applications \cite{Kim2009}.

Experimental investigations into spinodal decomposition in ternary alloys have been conducted for several systems, including Cu-Ni-Cr~\cite{findik1993sidebands}, Cu-Ni-Fe~\cite{LIVAK1974589}, Ti-Al-NB~\cite{rios2011spinodal}, Co-Ti-Fe~\cite{Singh1980}, Cu-Ni-Sn~\cite{spooner1980}, Fe-Cr-Mo~\cite{2002}, Al-Li-Zr~\cite{LouiseMakin1984}, Fe-Cr-Co~\cite{okada1978}, and In-Ga-As~\cite{chu1985}. However, these studies focus on ternary systems that exhibit a two-phase miscibility gap. Research has also been conducted on spinodal decomposition within two-phase miscibility gaps in metallic glasses, specifically in ternary systems such as Ni-Nb-Y~\cite{GOERIGK20093652, MATTERN2009903, MATTERN2010299}, Ge-Se-Ag~\cite{1Pattanayak}, as well as in higher order systems~\cite{CHANG20102483, SOHN201257, Park2012}. 

The foundational theoretical work by Cahn and Hilliard~\cite{jwcahn_jehilliard} established the basis for studying spinodal decomposition, introducing the Cahn-Hilliard equation to describe the temporal evolution of the composition field. Subsequently, Cahn~\cite{cahn1961spinodal} conducted analytical studies to predict the critical wavelength and the wavelength of maximum growth for composition fluctuations under the spinodal mechanism. De Fontaine~\cite{de1972analysis,de1973analysis,de1979configurational} extended the Cahn-Hilliard theory to multicomponent systems, examining the stability of multicomponent solid solutions in relation to diffusional processes and ordering. De Fontaine analyzed the properties of the bulk free energy function to predict a range of fluctuations that could grow via a spinodal mechanism. These findings were corroborated by Chen~\cite{CHEN1993683,CHEN19943503} and Eyre~\cite{eyrie}, who studied systems where $\alpha\beta$, $\beta\gamma$, and $\alpha\gamma$ interfaces are isotropic and possess identical interfacial energy. Bhattacharyya~\cite{bhattacharyya2003study} investigated spinodal decomposition in ternary systems, demonstrating that the relative values of the three interfacial energies can influence decomposition pathways. \textcolor{black}{More recent studies have further expanded this framework by incorporating complex interfacial anisotropies and nonlinear effects, providing deeper insight into the morphological evolution and pattern selection in multicomponent spinodal systems~\cite{ZHOU2021106349}.}

Early computer simulations of ternary spinodal decomposition leading to three-phase microstructures were predominantly conducted on 2D systems~\cite{CHEN1993683, CHEN19943503, bhattacharyya2003study, COPETTI200041, qiang, travasso, alfarraj}. However, 2D simulations in ternary systems often fail to capture critical features of the microstructures, such as co-continuous microstructures. The co-continuous network produced during the early stages breaks down, yielding isolated islands of the minority phase embedded within a sea of the majority matrix phase. Co-continuity may be retained well into the later stages if the volume fraction of the minority phase remains above 0.34. 

The early stages of decomposition in a ternary alloy were studied by examining the solutions to the linearized Cahn-Hilliard equations by  Morral and Cahn~\cite{morral1971spinodal}. Due to the extra compositional degrees of freedom in ternary and higher order systems, the ``decomposition tie line'' (the direction of composition fluctuations with the fastest growth rate in composition space) need not match with the tie-line containing the originally homogeneous alloy. This difference may arise from differences in gradient energy coefficients (i.e., interfacial energy) and/or atomic mobilities. Morral and Cahn derived analytical expressions for both the decomposition tie line and the wavelength of the fastest growth mode~\cite{morral1971spinodal}.

\textcolor{black}{The phase-field method has emerged as a powerful framework for modelling microstructure evolution owing to its ability to naturally capture interfaces at mesoscopic (nm–$\mu$m) length scales. It has been widely employed to study a broad spectrum of phenomena, including grain growth \cite{Chen199415752}, spinodal decomposition \cite{Cahn1961795,Cahn1962}, domain evolution \cite{Wang2004}, solidification \cite{Boettinger2002163,PhysRevE.57.4323}, recrystallization \cite{Chen2015}, Ostwald ripening \cite{Fan2002}, dislocation dynamics \cite{Rodney2003,Jin2001}, and fracture processes \cite{Henry2004,Spatschek2011}. Beyond these classical applications, phase-field formulations have been extended to a variety of complex phenomena, including cyclic degeneration in shape memory alloy single crystals \cite{XU2020105303}, fracture in composite systems \cite{ZHANG2019105008,LI2020105633}, electromigration-driven defect evolution in interconnects \cite{WU2025109792}, and thermodynamically consistent descriptions of fracture in temperature-dependent materials \cite{YIN2025110382}. Furthermore, the framework has proven particularly effective in capturing the influence of externally applied fields on microstructure evolution~\cite{Leo1998,Gururajan2007,Koyama2008,Mukherjee2016,Mukherjee20162}. Despite these successes, extending to three-dimensional multicomponent systems remains computationally demanding, especially for ternary systems that exhibit three-phase spinodal decomposition. As a result, substantial effort has focused on reducing computational cost through improved numerical algorithms and high-performance computing strategies~\cite{Muranushi_2012,vondrous2014parallel,miyoshi2017ultra,YANG2017133}. In parallel, artificial intelligence (AI) and machine learning (ML) methods have gained attention as promising tools for learning and accelerating microstructure evolution dynamics directly from physics-based simulation data~\cite{montes2021accelerating,hu2022accelerating,owaisprm,Ahmad_2025,TIWARI2025113518,gaikwad2025}. Nevertheless, most existing ML-assisted phase-field studies remain limited to binary or effectively two-phase systems, while three-dimensional ternary three-phase spinodal decomposition remains largely unexplored.}

This study aims to bridge this gap by developing an ML-enhanced phase-field framework specifically designed for ternary systems undergoing genuine three-phase spinodal decomposition in three dimensions. Our approach combines physics-based phase-field simulations, which accurately resolve early-stage instability and pattern selection, with an attention-enhanced ConvLSTM model trained to predict subsequent coarsening and morphological evolution. By learning from physically consistent simulations, the model captures long-time, topology-sensitive dynamics that would otherwise require prohibitively expensive full 3D computations. GPU-based phase-field solvers are employed as efficient, accessible tools to accelerate the generation of high-fidelity training data and baseline simulations. By tightly integrating physical modeling with data-driven temporal prediction in ternary three-phase microstructures, this work advances ML-assisted phase-field modeling beyond binary and two-dimensional studies and provides a scalable pathway for investigating complex multi-component alloy design problems.

\section{Phase-field model}
\label{pfmethod}
\begin{figure}
\centering
\subfigure[]{
  \includegraphics[scale=0.3]{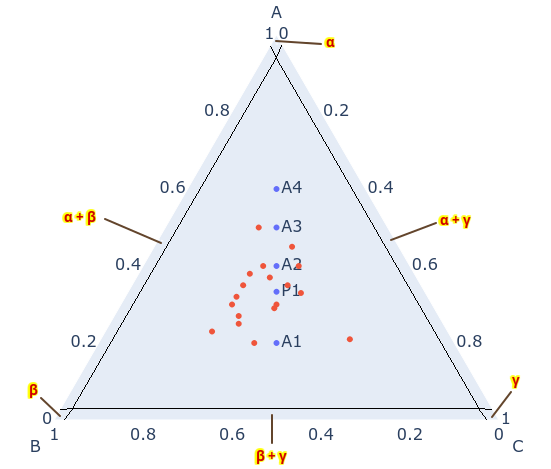}
}\\[1ex] 
\subfigure[]{
  \includegraphics[scale=1.5]{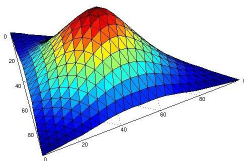}
}
\caption{(a) Schematic of a ternary phase diagram for a system with $\chi_{AB} = \chi_{BC} = \chi_{AC} > 2.0$. The red circle indicates the training compositions, while the blue circles on the phase diagram represent the predicted compositions [see Table~\ref{tabcomp}]. (b) Schematic diagram of free energy as a function of alloy composition.}
\label{spinodal}
\end{figure}

\subsection{Formulation for ternary system}
The present study employs a phase-field model to generate a training dataset of microstructure evolution during the spinodal decomposition. We employ a ternary version of the Cahn-Hilliard model for a system with components $A-B-C$~\cite{bhattacharyya2003study,lee2012practically, ghosh2017particles}. The (bulk) free energy per atom, $f(c_{A},c_{B},c_{C})$, of a homogeneous solid solution with a composition $(c_{A}, c_{B}, c_{C})$, is given by~\cite{jwcahn_jehilliard},
\begin{equation}
f\left(c_A,c_B,c_C\right)=  \frac{1}{2} \sum_{i \neq j}\chi_{ij}c_{i}c_{j}+\sum c_{i} \ln c_{i}.
\label{reg_sol_ter}
\end{equation}
In the above equation, $i$ and $j$ refer to one of the three components $\left\{ A, B, C \right\}$, $c_{i}$ is the mole fraction of component $i$, and $\chi_{ij}$ is the interaction parameter that sets the barrier height between the free energy densities of A-rich $\alpha$, B-rich $\beta$  and C-rich $\gamma$ phases. The (temperature-normalized) regular solution parameter $\chi_{ij}$ for the binary $i-j$ system is defined by,
\begin{equation}
\chi_{ij}=\frac{Z[2E_{ij}-E_{ii}-E_{jj}]}{2k_{B}T}.
\label{kai_eq}
\end{equation}
In the above equation, $E_{ij}$ is the energy of the bond between nearest neighbor pairs $ij$, $Z$ is the number of bonds per atom, $k_B$ is the Boltzmann constant, and $T$ is the absolute temperature. 

\begin{table}[!htb]
 \begin{center}
 \begin{tabular} { | p {1.5 cm} || p {1.5 cm} | p {1.5 cm} | p {1.5 cm} |}
 \hline
 \multicolumn{4} { |c| }{Composition of predicted microstructures} \\
 \hline
Alloys & $C_A$ & $C_B$ & $C_C$ \\
 \hline
A1   & 0.200    & 0.400 & 0.400 \\
 & & &\\
P1  & 0.333   & 0.333   & 0.333 \\
& & &\\
A2  & 0.400  & 0.300  &  0.300 \\
& & &\\
A3  & 0.500  & 0.250  &  0.250 \\
& & &\\
A4  & 0.600  & 0.200  &  0.200 \\
\hline
\end{tabular}
\end{center}
\caption{Alloy compositions of predicted microstructures. These compositions are not a part of the training set. The corresponding phase diagram is shown in Figure~\ref{spinodal}.}
\label{tabcomp}
\end{table}

 The binary $AB$ system exhibits a miscibility gap when $\chi_{AB}$ exceeds a critical value, $\chi_{c} = 2.0$. In our study, we are interested in a ternary miscibility gap, where three phases (terminal solutions) are in equilibrium. All the interaction parameters $(\chi_{AB}, \chi_{BC}, \chi_{AC})$ in our model are greater than the critical value $\chi_{c}$. Figure~\ref{spinodal} shows a schematic of an isothermal section of the phase diagram, having a three-phase miscibility gap in the middle of the Gibbs triangle. This figure also shows a schematic of the free energy per atom as a function of alloy composition. The free energy function has three minima and one maximum. Alloys whose compositions fall near the middle of this ternary miscibility gap may be homogenized by heating them to a high temperature (at which $\chi_{ij} < \chi_c)$. Upon quenching, they would undergo phase separation if the composition falls in a region where at least one of the principal curvatures of the free energy surface is negative. This phase separation process may be initiated by a spinodal mechanism.

The total Helmholtz free energy $F$, for an isotropic, ternary system with compositional inhomogeneities, is expressed in terms of the following Cahn-Hilliard free energy functional:
\begin{equation}
\frac{F}{k_{B}T} = N_v\int_v\left[f(c_A,c_B,c_C) + \sum_{i \in A,B,C} {\kappa}_{i}({\nabla c_i})^{2}\right]dV ,
\label{ter_cahn}
\end{equation}    
where $N_{v}$ is the number of atoms per unit volume (assumed to be independent of composition and position), and ${\kappa}_{i}$ is the gradient energy coefficient associated with a gradient of species composition $c_{i}$.

To track the temporal evolution of respective composition fields, a ternary Cahn-Hilliard equation can be derived~\cite{bhattacharyya2003study} using the Continuity equation, results of Kramer~\cite{kramer1984interdiffusion}, Gibbs-Duhem equations, and Onsager relations,
\begin{eqnarray}
\frac{\partial c_A}{\partial t} = M_{AA}[\nabla^2(\frac{\partial f}{\partial c_A})-2(\kappa^{AA})\nabla^4 c_A-2\kappa^{AB}\nabla^4 c_B]\nonumber \\
 + M_{AB}[\nabla^2(\frac{\partial f}{\partial c_B})-2(\kappa^{BB})\nabla^4 c_B-2\kappa^{AB}\nabla^4 c_A],
\label{ternary-equation_1}
\end{eqnarray}
\begin{eqnarray}
\frac{\partial c_B}{\partial t}  = M_{BB}[\nabla^2(\frac{\partial f}{\partial c_B})-2(\kappa^{BB})\nabla^4 c_B-2\kappa^{AB}\nabla^4 c_A]\nonumber \\
 + M_{AB}[\nabla^2(\frac{\partial f}{\partial c_A})-2(\kappa^{AA})\nabla^4 c_A-2\kappa^{AB}\nabla^4 c_B].
\label{ternary-equation_2}
\end{eqnarray}
Effective mobilities $M_{AA}$, $M_{BB}$ and $M_{AB}=M_{BA}$ (which are the elements of a symmetric $2 \times 2$ mobility matrix) are as follows:
\begin{eqnarray}
M_{AA} & = &(1-c_A)^2M_A+c_A^2(M_B+M_C),\nonumber \\
M_{BB} & = &(1-c_B)^2M_B+c_B^2(M_A+M_C),\nonumber \\
M_{AB} & = &M_{BA}  = c_Ac_BM_C  - (1-c_A)c_BM_A\nonumber \\ & & + c_AM_B(1-c_B).
\label{mobilitiese}
\end{eqnarray}
In writing these equations, we have used the following definitions: 
$\kappa^{AA}=\kappa_A+\kappa_C$, $\kappa^{BB}=\kappa_B+\kappa_C$ and $\kappa^{AB}=\kappa^{BA}=\kappa_C$, which form a 2$\times$2 matrix of gradient coefficients in ternary systems. The following $\kappa$ matrix is used in our study.
\begin{equation}
    \begin{pmatrix} \kappa^{AA} & \kappa^{AB}\\ \kappa^{BA} & \kappa^{BB} \end{pmatrix}=\begin{pmatrix} 5 & 2.5\\ 2.5 & 5 \end{pmatrix}.
\end{equation}
All the parameter values used for phase-field simulation are given in Table~\ref{tab:pfparams}. All values are dimensionless.
\begin{table}[!htb]
\centering
\begin{tabular*}{\columnwidth}{|p{4cm}|@{\extracolsep{\fill}}p{4cm}|}
\hline
\multicolumn{2}{|c|}{Phase-field simulation parameters} \\
\hline
Parameter & Value \\
\hline
$n_x, n_y$ (2D) & $128,\,128$ \\
$n_x, n_y, n_z$ (3D) & $128,\,128,\,128$ \\
$dx, dy, dz$ & $1.0,\,1.0,\,1.0$ \\
$M_{AA},\,M_{BB},\,M_{AB}$ & $1.0,\,1.0,\,0.5$ \\
$\chi_{AB},\,\chi_{BC},\,\chi_{AC}$ & $5.0,\,5.0,\,5.0$ \\
$\kappa_A,\,\kappa_B,\,\kappa_C$ & $2.5,\,2.5,\,2.5$ \\
$dt$ (2D) & $0.0005$ \\
$dt$ (3D) & $0.0001$ \\
\hline
\end{tabular*}
\caption{Phase-field simulation parameters used for ternary spinodal decomposition simulations. Different time steps are used in 2D and 3D for numerical stability.}
\label{tab:pfparams}
\end{table}

\subsection{Numerical implementation}
\label{numerical}
The semi-implicit Fourier spectral method, originally developed by Shen and Chen~\cite{CHEN1998147} for solving the binary Cahn-Hilliard equation, can be readily extended to solve the ternary Cahn-Hilliard equations. Fourier transform of Eq.~\ref{ternary-equation_1} and~\ref{ternary-equation_2} leads to: 
\begin{align}
\frac{\partial\tilde{c}_{A}(\textbf{k},t)}{\partial t} 
&= M_{AA}\left[-k^2\tilde{g}_{A}(\textbf{k}) - 2\kappa^{AA}k^4\tilde{c}_{A} - 2\kappa^{AB}k^4\tilde{c}_{B} \right] \nonumber \\
&\quad + M_{AB}\left[-k^2\tilde{g}_{B}(\textbf{k}) - 2\kappa^{AB}k^4\tilde{c}_{A} - 2\kappa^{BB}k^4\tilde{c}_{B} \right],
\label{CH_fou_final1} \\
\frac{\partial\tilde{c}_{B}(\textbf{k},t)}{\partial t}  
&= M_{BB}\left[-k^2\tilde{g}_{B}(\textbf{k}) - 2\kappa^{BB}k^4\tilde{c}_{B} - 2\kappa^{AB}k^4\tilde{c}_{A} \right] \nonumber \\
&\quad + M_{AB}\left[-k^2\tilde{g}_{A}(\textbf{k}) - 2\kappa^{AB}k^4\tilde{c}_{B} - 2\kappa^{AA}k^4\tilde{c}_{A} \right],
\label{CH_fou_final}
\end{align}
where $ g_A  = \frac{\partial f}{\partial c_A}$, $ g_B  = \frac{\partial f}{\partial c_B}$, $\textbf{k}$ is a vector in Fourier space and $k=\mid\textbf{k}\mid $. $ \tilde{c}_{A}(\textbf{k},t)$ and $ \tilde{c}_{B}(\textbf{k},t)$ are the Fourier transforms of the respective composition fields in real space. Using forward  difference to approximate $\partial \tilde{c}_{X} / \partial t$, we get:
\begin{eqnarray}
\frac{\partial \tilde{c}_{X}}{\partial t} & = &\frac{\tilde{c}_{X}(\textbf{k},t+\Delta t) - \tilde{c}_{X} (\textbf{k},t)} {\Delta t},
\end{eqnarray}
where $X \in { A,B }$. Treating the linear terms, $\tilde{c}_{A}$ and $\tilde{c}_{B}$ implicitly, and the non-linear terms, $\tilde{g}_{A}$ and $\tilde{g}_{B}$ explicitly, we arrive at the following equations in our semi-implicit formulation: 
\begin{eqnarray}
\left[                  
\begin{array}{rl}       
Q_{11}& Q_{12}\\
Q_{21}& Q_{22}\\
\end{array}
\right]
\left[
\begin{array}{c}
\tilde{c}_{A}(t+\Delta t)\\
\tilde{c}_{B}(t+\Delta t)\\
\end{array}
\right]
= \nonumber
\\
\left[
\begin{array}{c}
\tilde{c}_{A}(t)- M_{AA} k^{2}\tilde{g}_{A}\Delta t - M_{AB} k^{2} \tilde{g}_{B} \Delta t\\
\tilde{c}_{B}(t)- M_{BB} k^{2} \tilde{g}_{B} \Delta t - M_{AB} k^{2} \tilde{g}_{A} \Delta t\\
\end{array}
\right],
\end{eqnarray}
where,
\begin{eqnarray}
Q_{11} & = & 1+2M_{AA}\kappa^{AA}k^{4}\Delta t+2M_{AB}\kappa^{AB}k^{4}\Delta t, \nonumber \\ 
Q_{12} & = & 2(M_{AA}\kappa^{AB}+M_{AB}\kappa^{BB})k^{4}\Delta t, \nonumber \\
Q_{21} & = & 2(M_{BB}\kappa^{AB}+M_{AB}\kappa^{AA})k^{4}\Delta t, \nonumber \\
Q_{22} & = & 1+2M_{BB}\kappa^{BB}k^{4}\Delta t+2M_{AB}\kappa^{AB}k^{4}\Delta t.
\end{eqnarray}
Given the composition fields at time $t$, a numerical solution of these equations yields the composition fields at $t+\Delta t$. Repeating these steps constitutes a computer simulation.

Fourier transform renders Eq.~\ref{CH_fou_final1} and~\ref{CH_fou_final}  into an ODE, from the original PDE in Eq.~\ref{ternary-equation_1} and~\ref{ternary-equation_2}, which now has to be time integrated to complete the numerical solution process. As $g(c)$ is a non-linear function of $c$, and there being no simple expression relating the Fourier transform of $c$ to the Fourier transform of $g(c)$, we evaluate $g(c)$ in the real space at the current instant of time ($t$) and compute its Fourier transform to obtain $\tilde{g_c}(\textbf{k})$, while all the other terms in Eq.~\ref{CH_fou_final1} and~\ref{CH_fou_final} are evaluated at the future instant of time $t+\Delta t$.


\begin{figure}
\centering
\includegraphics[scale=0.37]{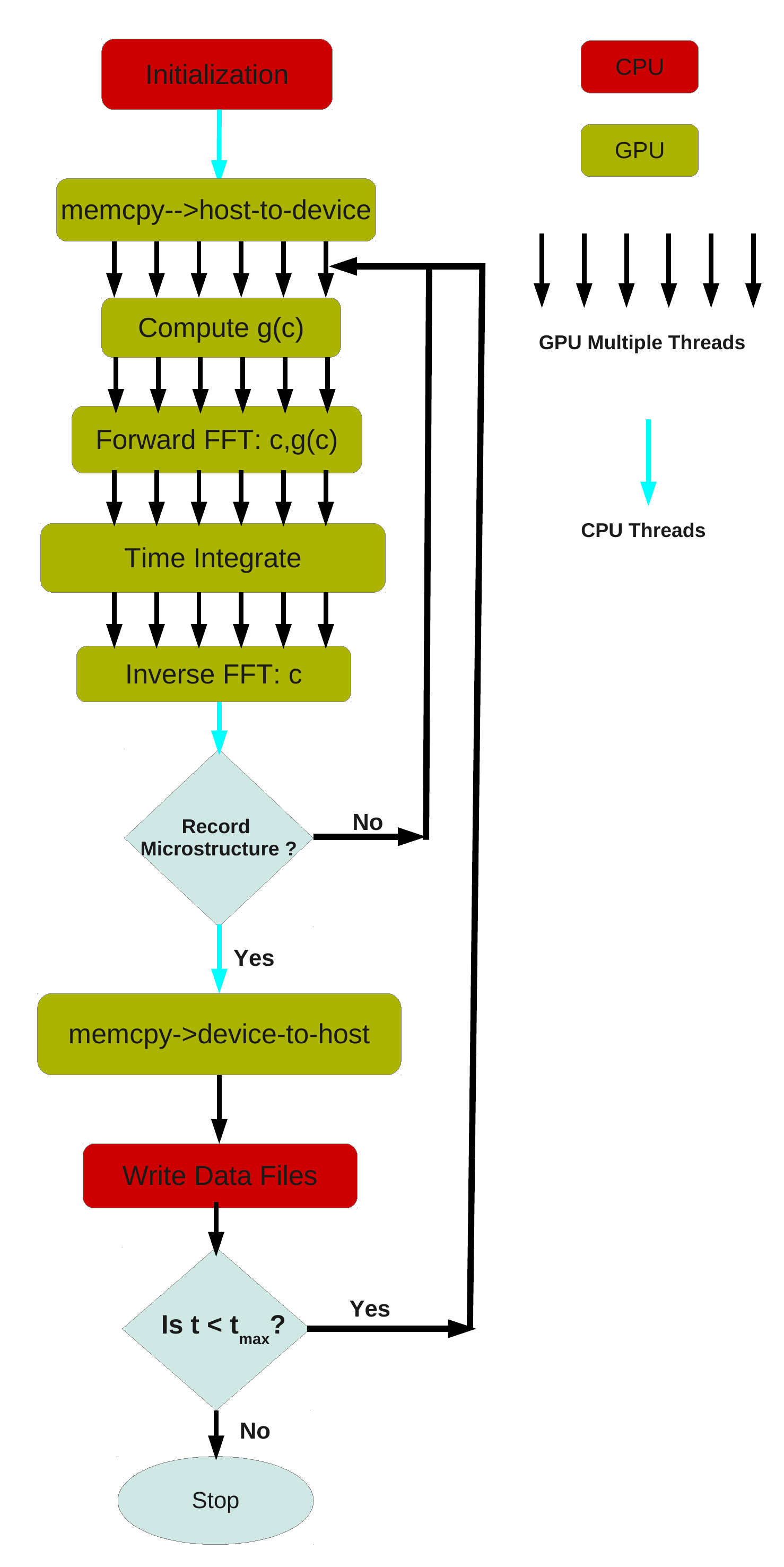}
\caption{Flowchart depicting the algorithm with clearly demarcated serial and parallel steps executed on the CPU and the GPU, respectively. The current timestep and the maximum number of time steps are denoted by $t$ and $t_{max}$, respectively.}
\label{flowchart}
\end{figure}

All simulations and benchmarks reported in this work are executed on a CPU-based server and a GPU-accelerated workstation. The CPU platform consists of an Intel Xeon Gold 6226R processor (16 physical cores, base clock 2.90 GHz) with 32 GB of host RAM. The accelerator platform is an NVIDIA RTX A5000 GPU with 24 GB of device memory. The nominal base clock of the GPU is reported as 1.170 GHz (actual runtime frequency may vary due to boost behavior). The measured peak memory bandwidths are 140.7 GB/s for the CPU and 768.0 GB/s for the GPU, and the RTX A5000 exposes 8192 CUDA cores, enabling massive thread-level parallelism.

The numerical algorithm is implemented in CUDA C. Serial control logic and one-time initialization are performed on the CPU, while the computationally intensive, data-parallel components of the solver are executed on the GPU. The overall workflow is illustrated in Figure~\ref{flowchart}. In particular, the semi-implicit spectral solution of the Cahn--Hilliard equations is executed on the GPU. The Fourier and inverse Fourier transforms at each timestep are computed using the optimized cuFFT library provided for NVIDIA hardware. Since the major computations are performed directly on the GPU, the composition fields remain in GPU memory throughout time integration, reducing the need for frequent data transfers between the CPU and GPU and improving overall efficiency. Although the CUDA-based implementation limits portability to NVIDIA GPUs, it enables substantial acceleration for the present problem. Other details are provided in Section~S1, Supporting Information (SI).


\begin{figure}
\centering
\subfigure[$P1, t=20,100,300$]{
\includegraphics[scale=0.28]{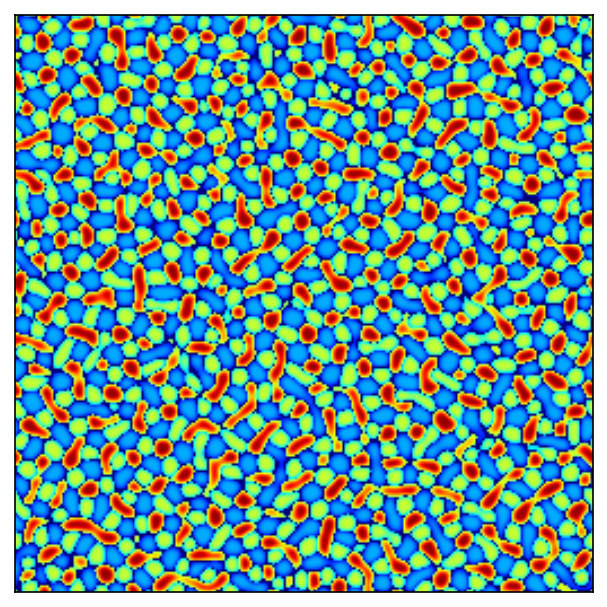}\label{2a}
\includegraphics[scale=0.28]{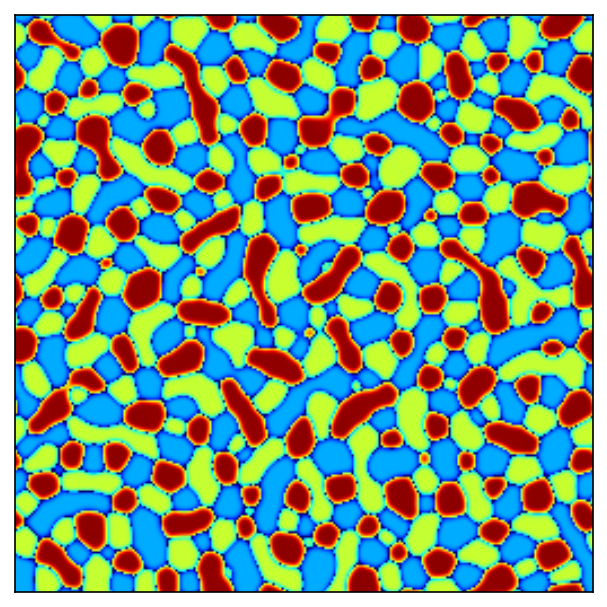}
\includegraphics[scale=0.28]{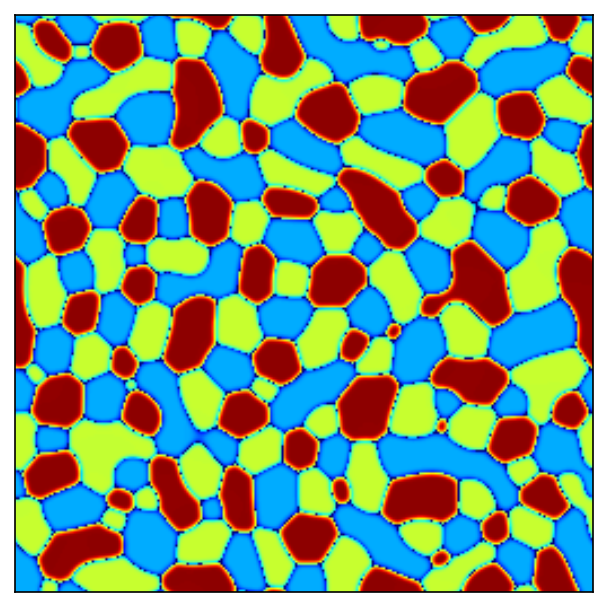}}\\
\subfigure[$A1, t=11,25,300$]{
\includegraphics[scale=0.28]{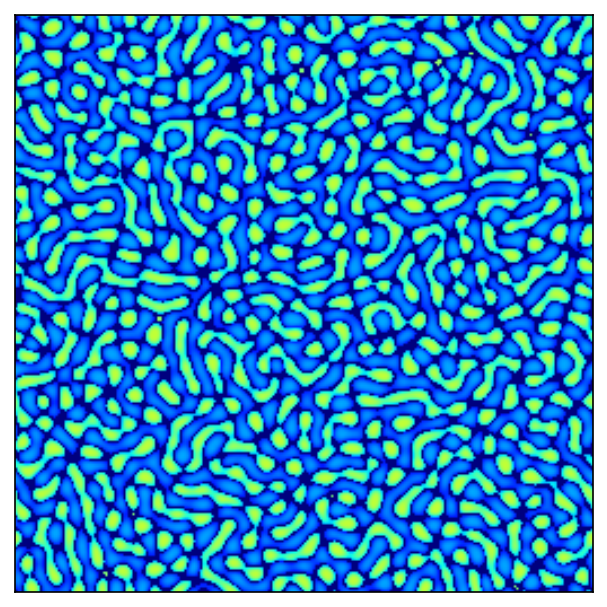}\label{2b}
\includegraphics[scale=0.28]{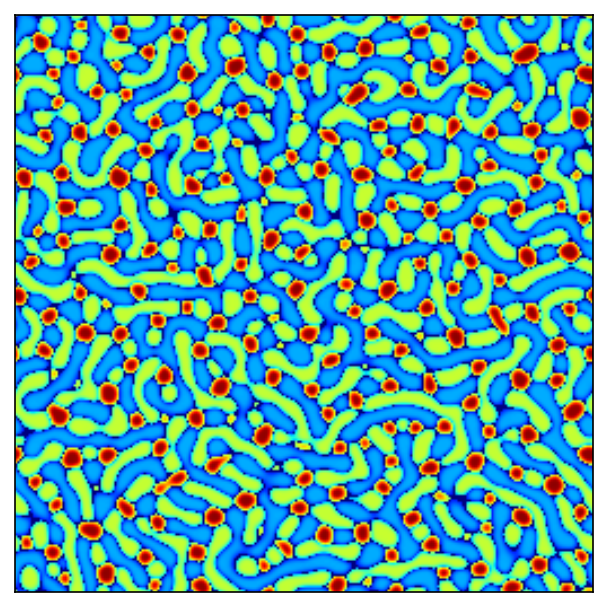}
\includegraphics[scale=0.28]{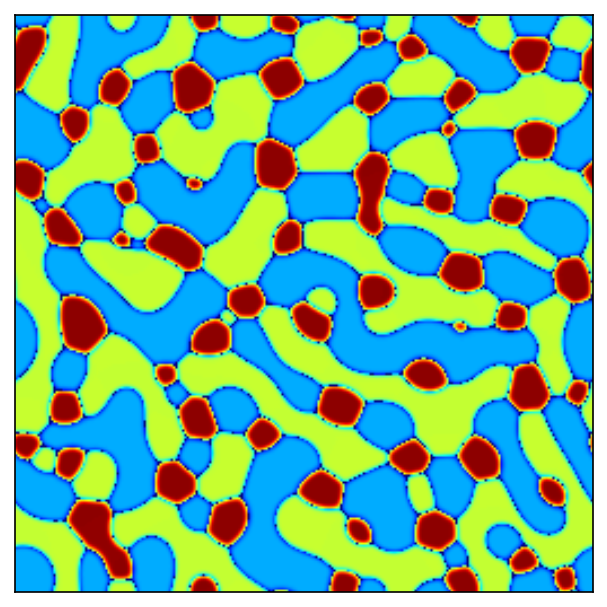}}\\ 
\subfigure[$A3, t=10,20,300$]{
\includegraphics[scale=0.28]{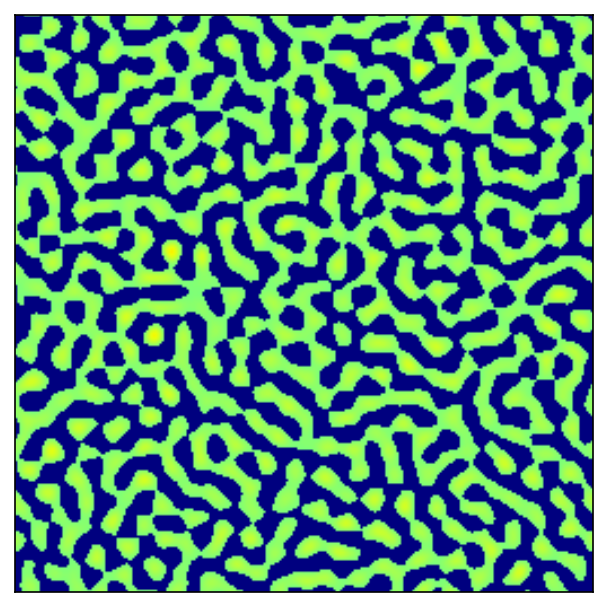}\label{2c}
\includegraphics[scale=0.28]{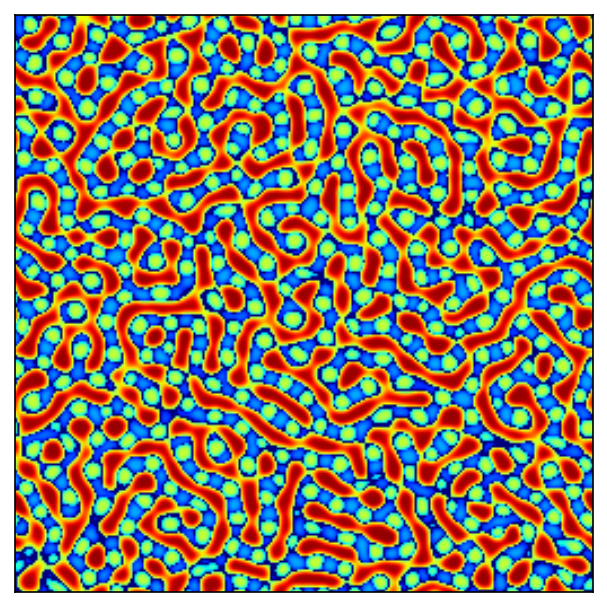}
\includegraphics[scale=0.28]{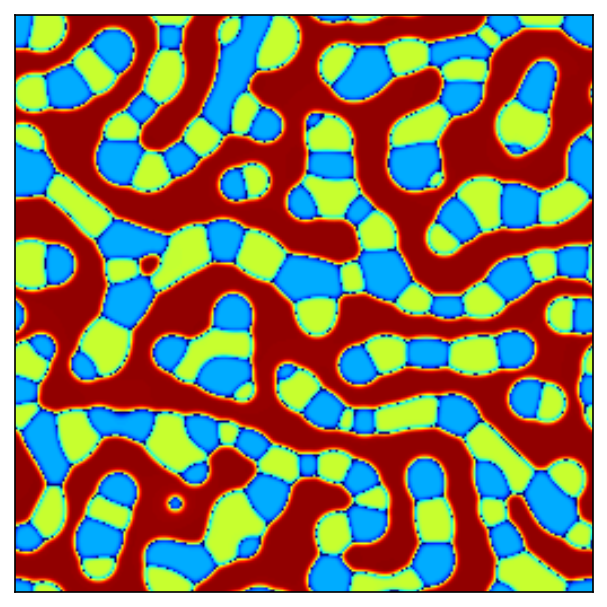}}
\caption{Time evolution of microstructures (scalar field maps) of alloys (a) P1, (b) A1, and (c) A3 in a simulation box of $256\times256$ grid points. We plot a scalar field that combines $c_A, c_B$, and $c_C$ values to improve contrast. Corresponding composition plots are provided in Figures S1-S3, Supporting Information (SI). The temporal evolution videos are also provided as supplementary video files. Alloys A1 and A3 show the two-stage spinodal decomposition, classified as Type I and Type II, respectively. The time steps for the plots are carefully chosen to effectively convey the two-stage mechanism across different compositions.}
\label{bin_2D}
\end{figure}

\section{Microstructure evolution in 2D and 3D}
\label{discussion}

\begin{figure}
\centering
\subfigure[$P1, t=23, 33, 50$]{
\includegraphics[scale=0.2]{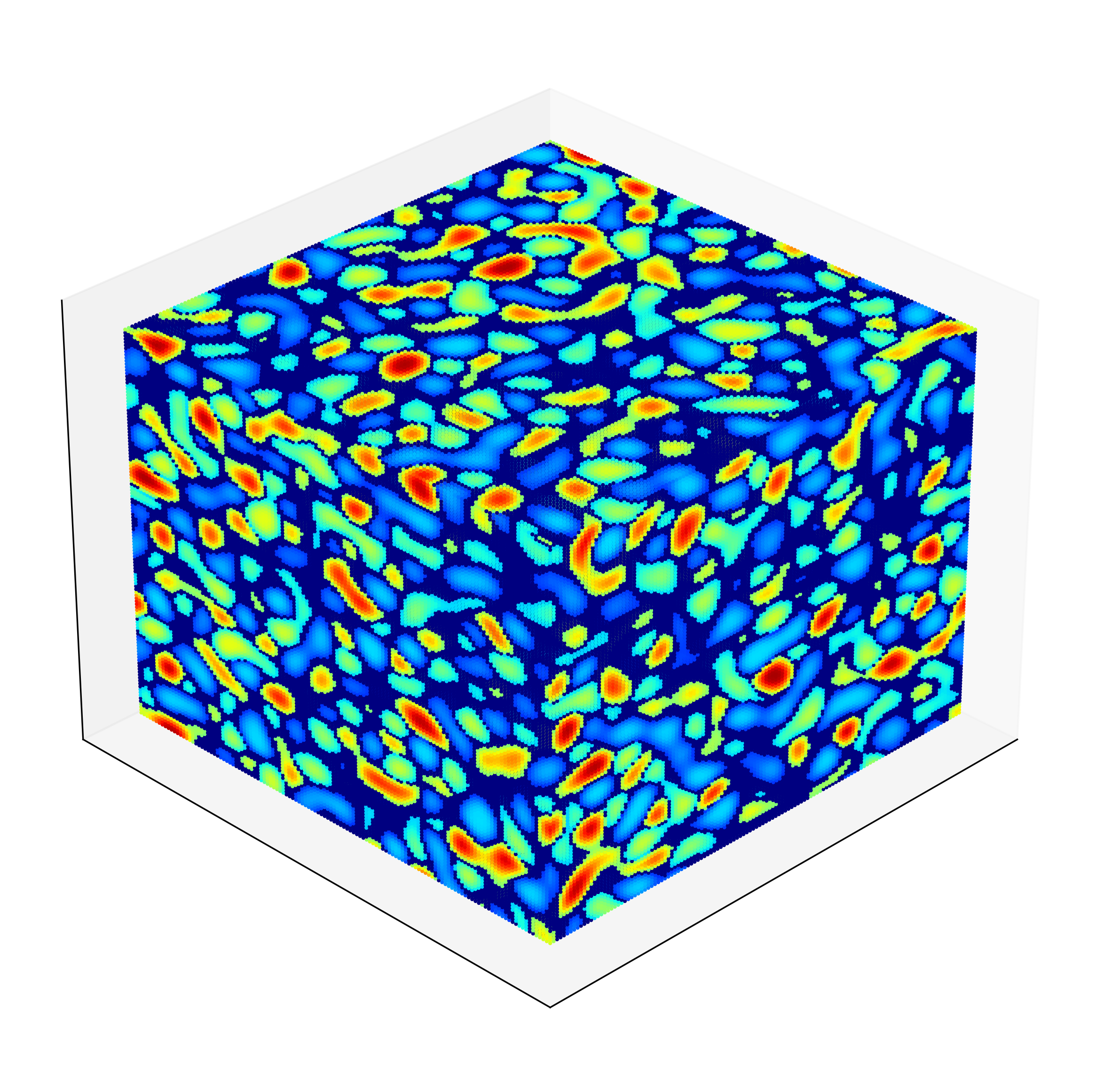}
\includegraphics[scale=0.2]{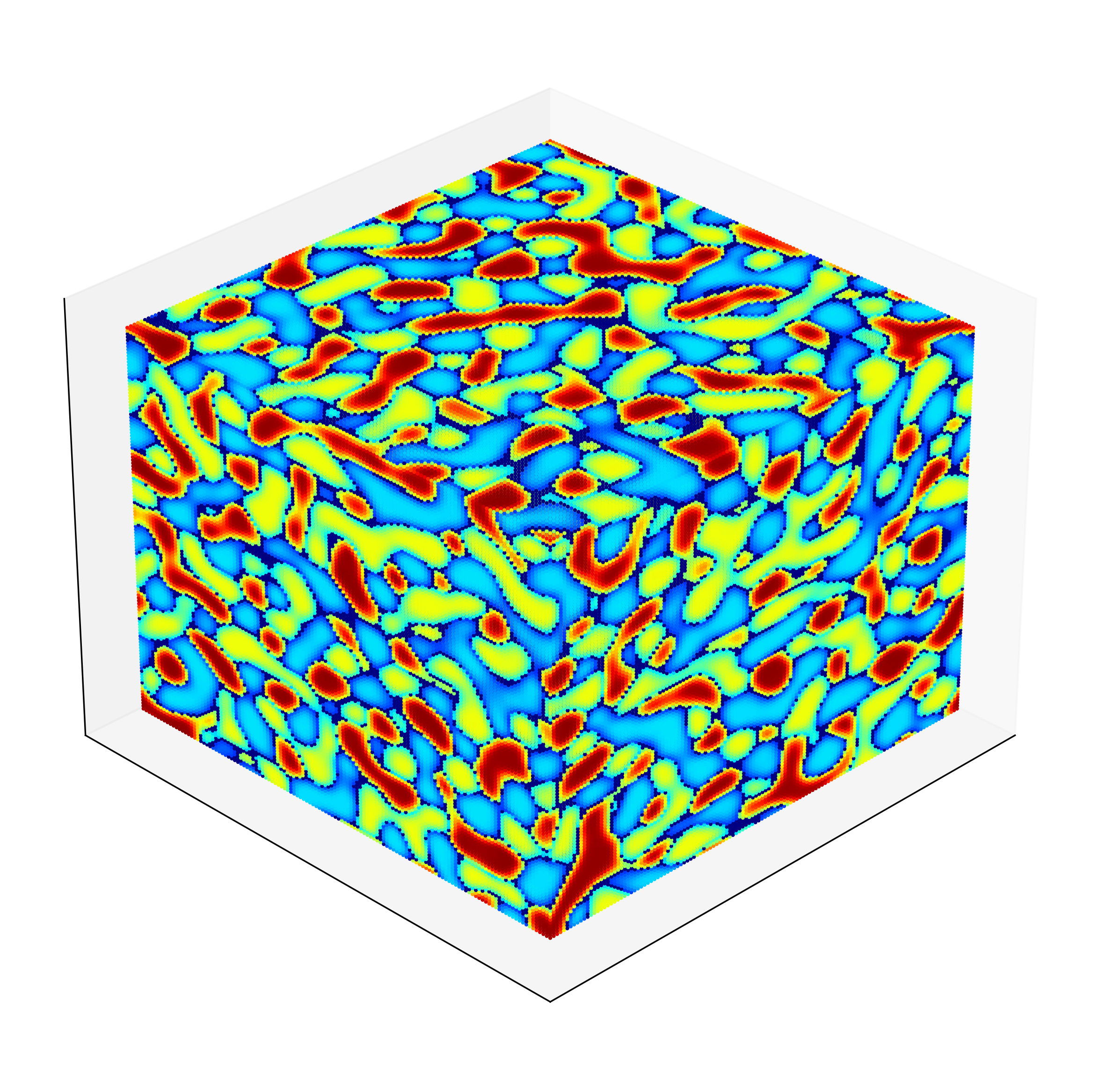} 
\includegraphics[scale=0.2]{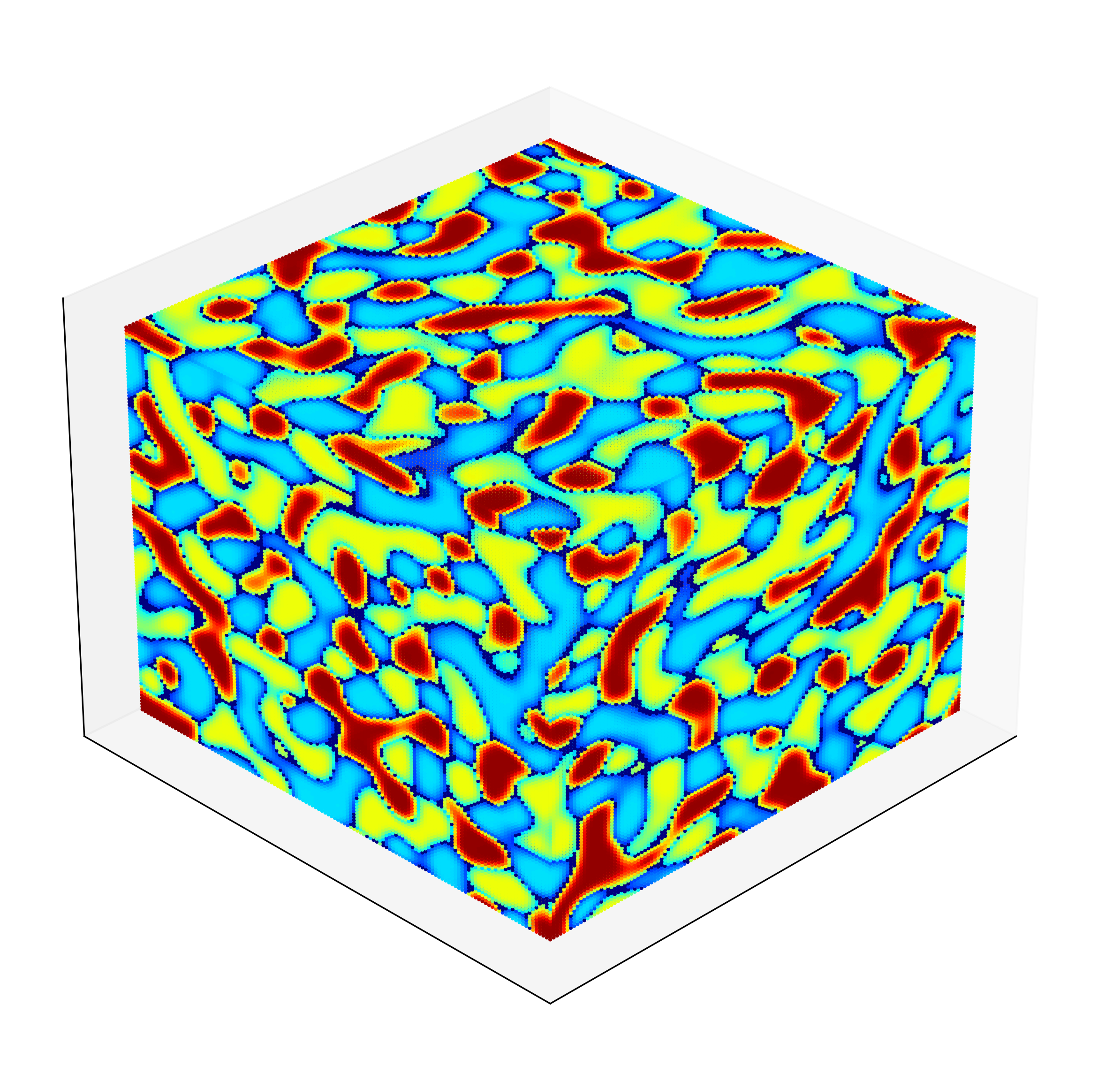}\label{3a}} \\
\subfigure[$A1, t=17, 21, 50$]{
\includegraphics[scale=0.2]{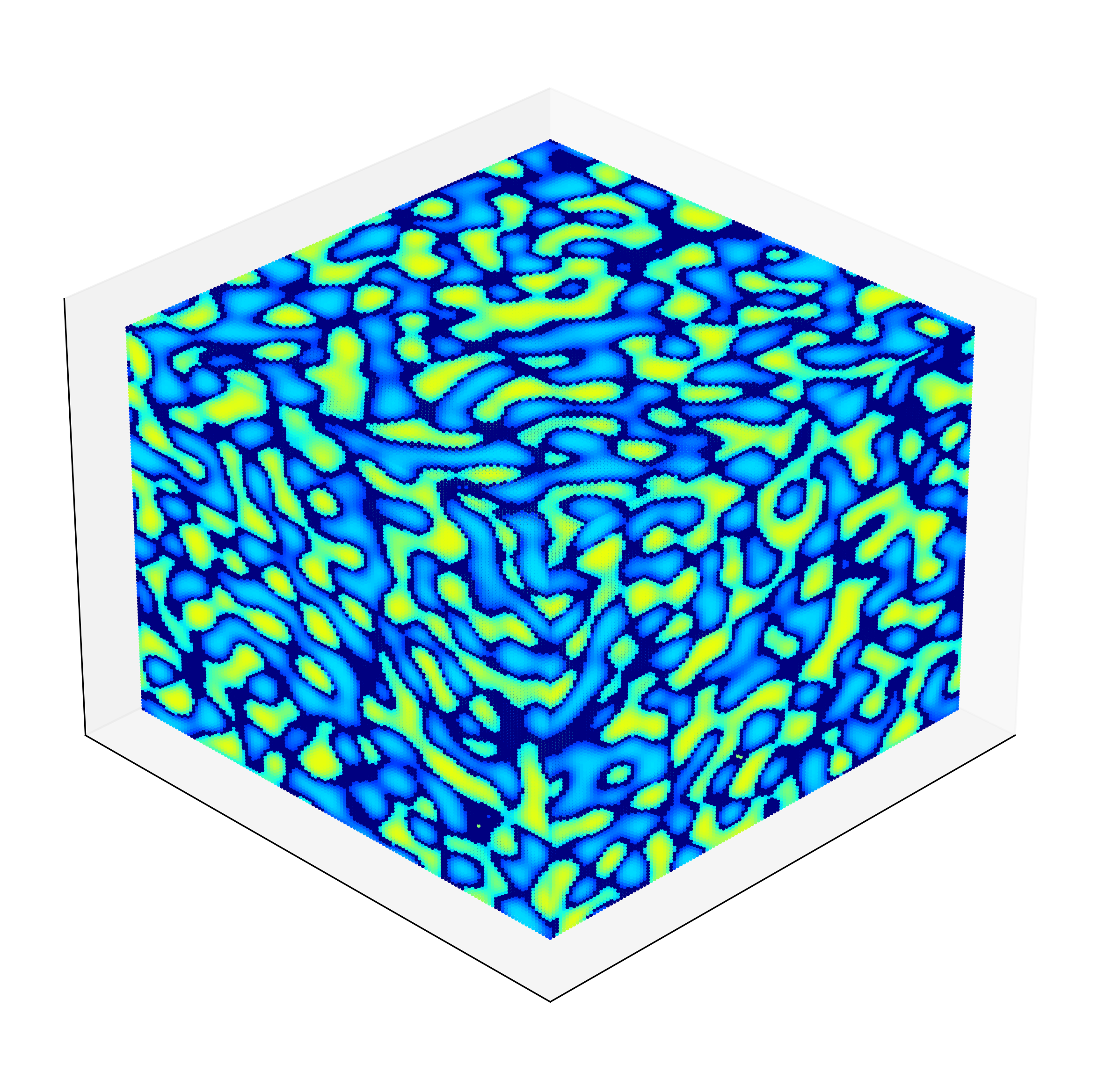}
\includegraphics[scale=0.2]{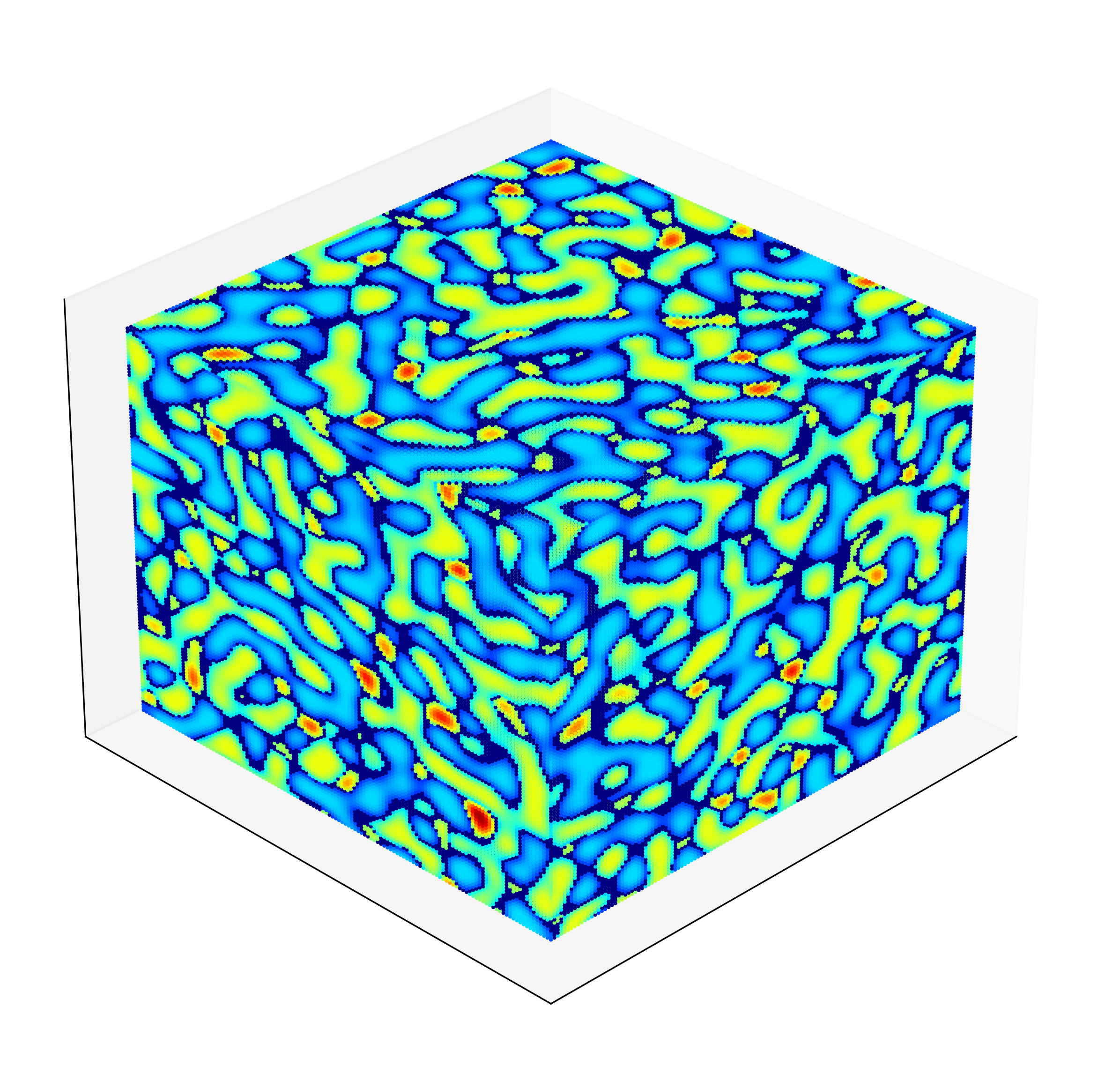} 
\includegraphics[scale=0.2]{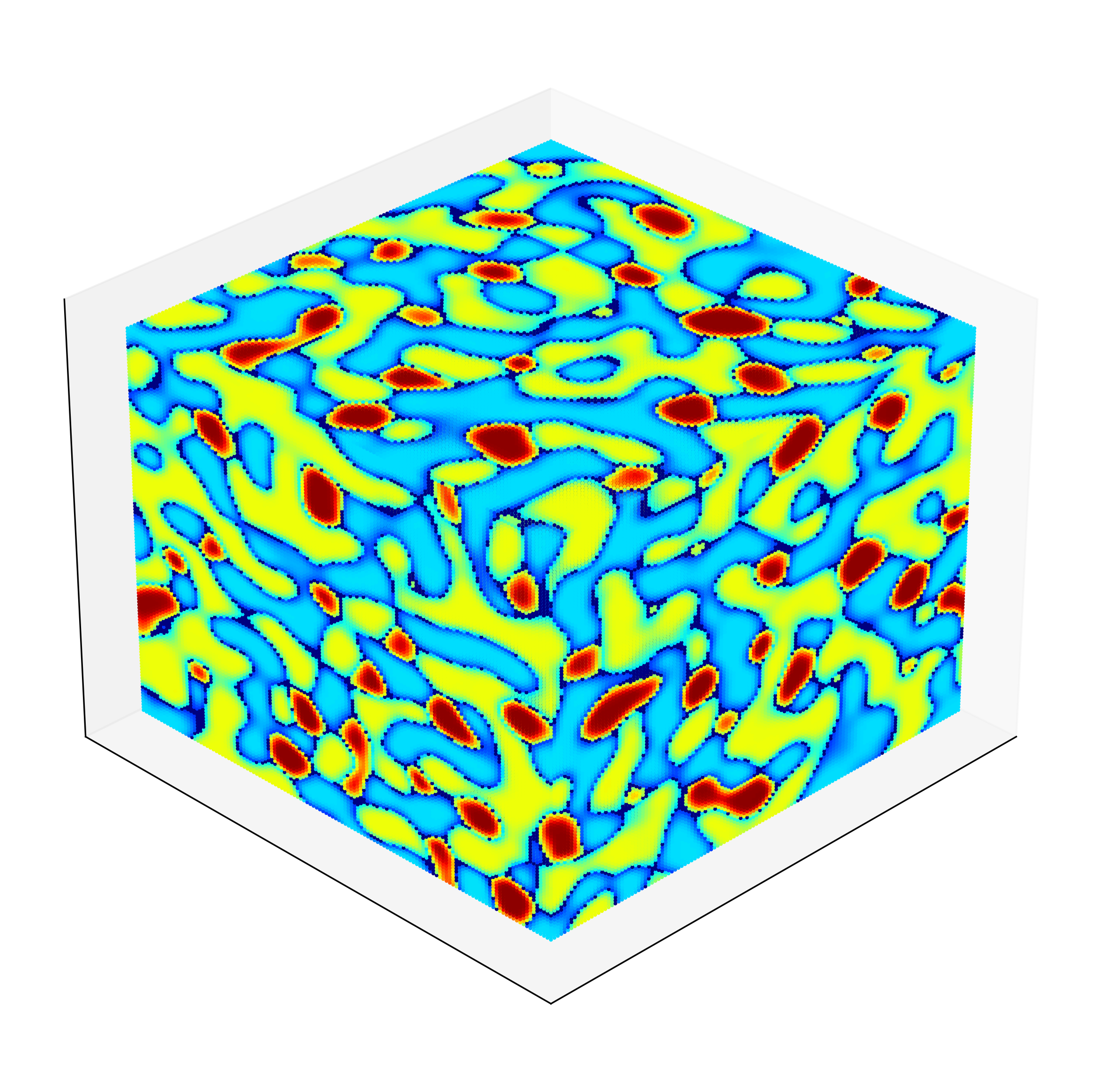}\label{3b}} \\ 
\subfigure[$A3, t=12, 32, 50$]{
\includegraphics[scale=0.2]{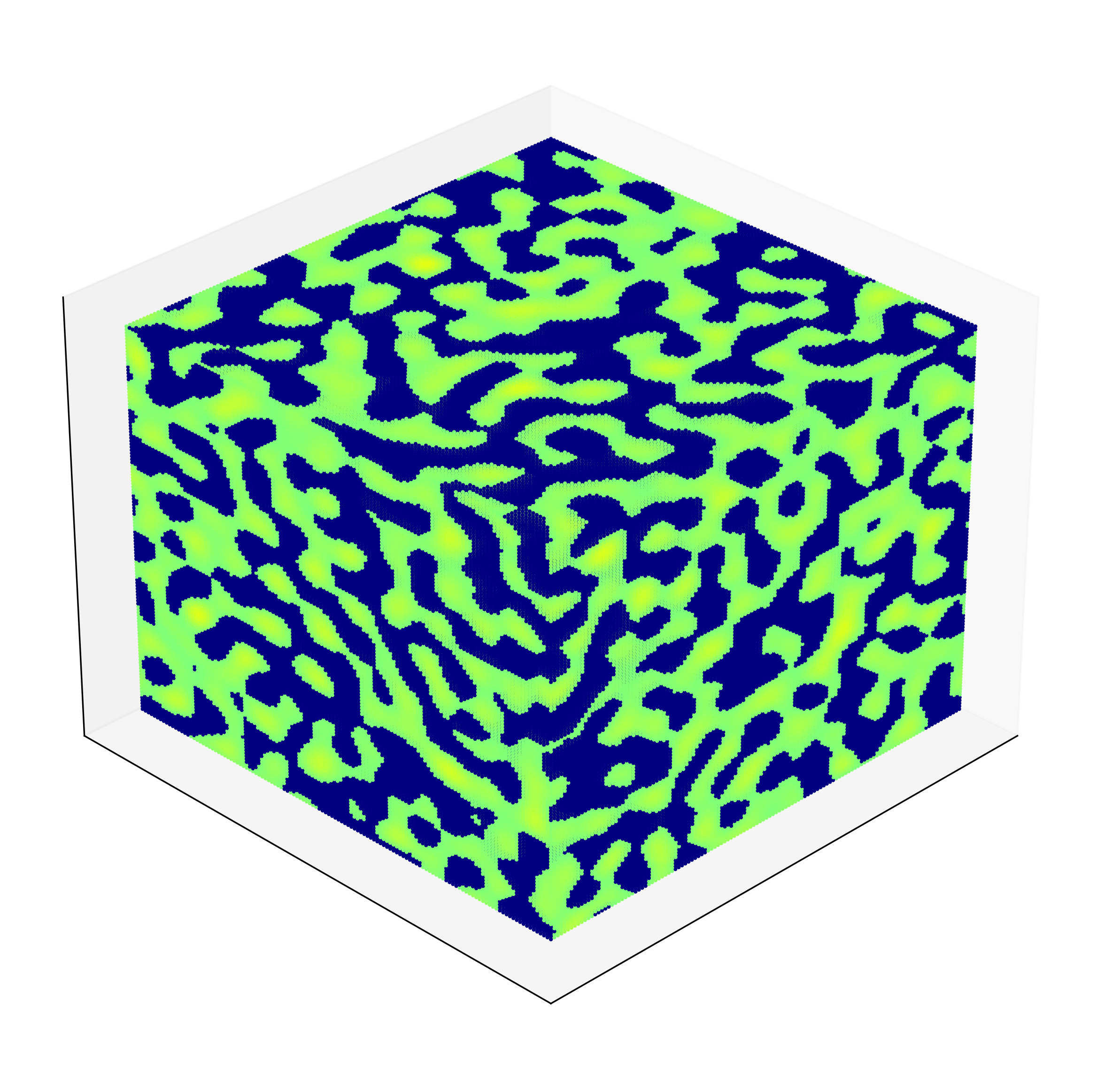}
\includegraphics[scale=0.2]{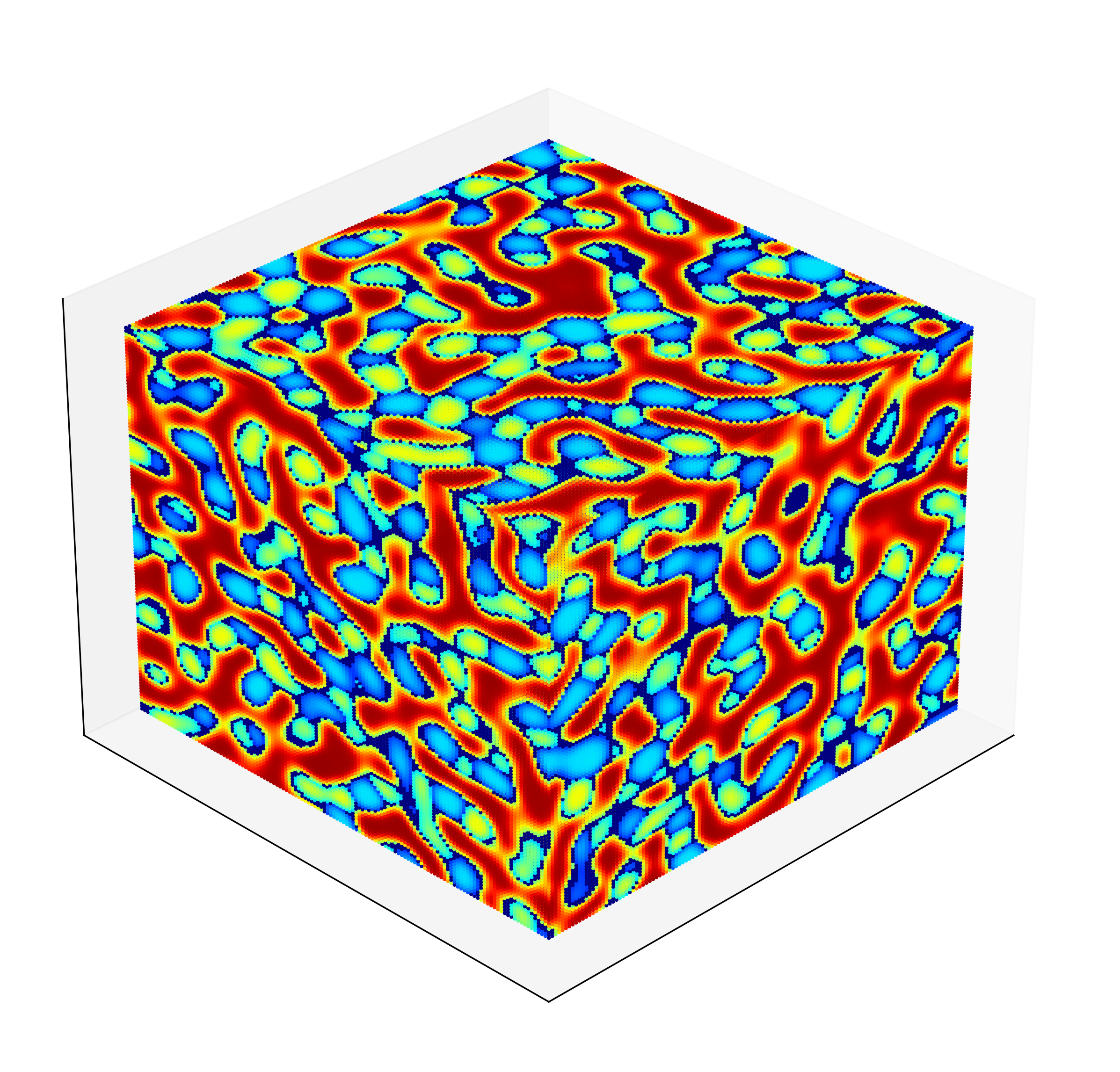} 
\includegraphics[scale=0.2]{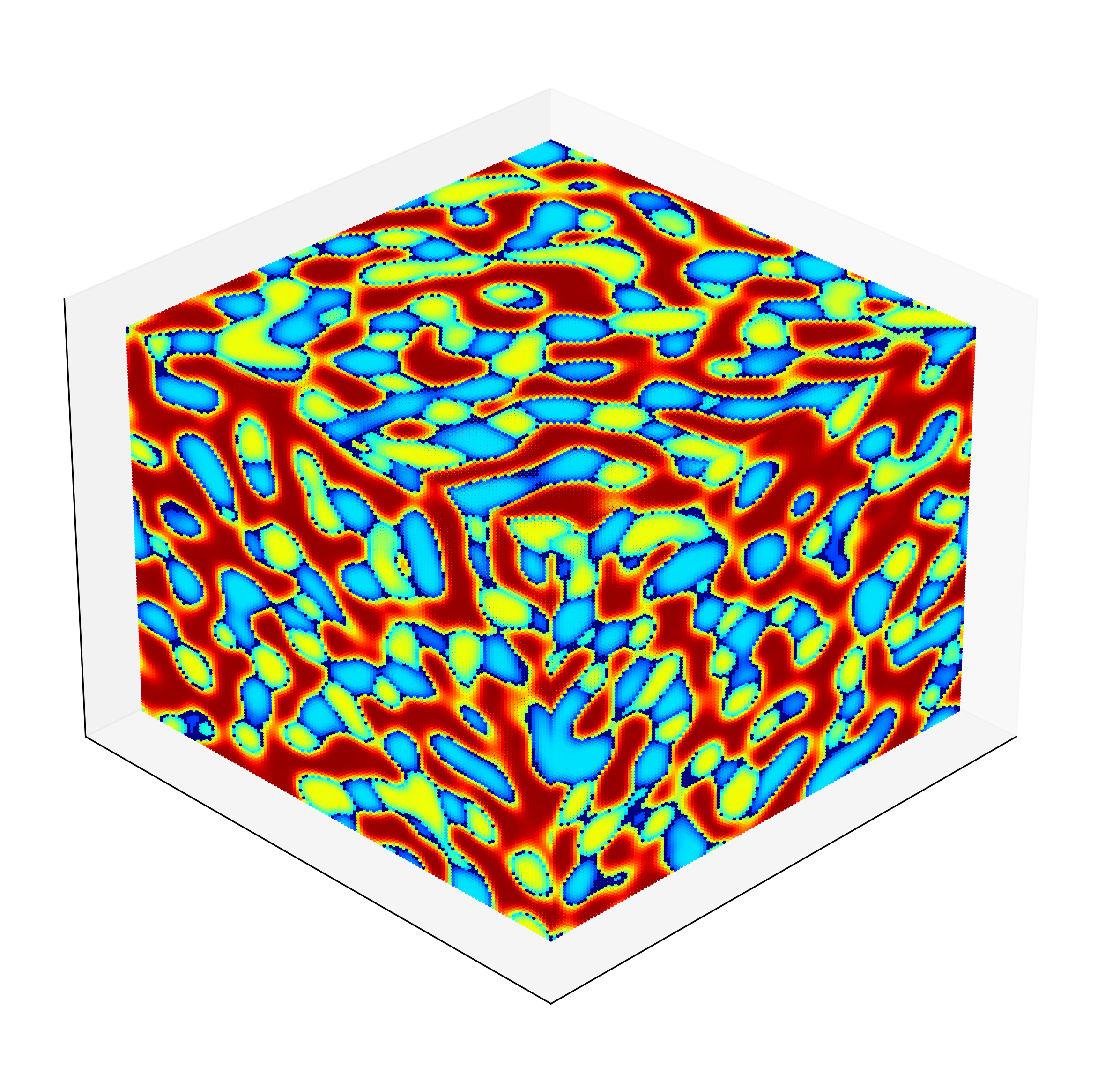}\label{3c}}
\caption{Time evolution of microstructures (scalar field maps) of alloys P1, A1, and A3 in a simulation box of $128\times128\times128$ grid points. Observe the similarity in mechanisms with 2D (Figure~\ref{bin_2D}). Microstructure evolutions of all the compositions are shown in Figures S4-S5, SI. The time steps for the plots are carefully chosen to effectively convey the two-stage mechanism across different compositions.}
\label{maps}
\end{figure}

Among the five prediction compositions [Figure~\ref{spinodal} and Table~\ref{tabcomp}], we select three alloys in the ternary phase diagram and present a detailed analysis of microstructure evolution. 
All the (scaled) regular solution parameters for the three ternary systems are taken to be equal and positive: $(\chi_{AB}, \chi_{BC}, \chi_{AC}) = (5.0, 5.0, 5.0)$, ensuring a ternary miscibility gap. All the alloys are in a compositionally homogeneous state at the start of the simulation, and they are assumed to be quenched to a temperature at which the phase diagram exhibits a ternary miscibility gap, resulting in a three-phase equilibrium.
We consider the phase separation behavior in these alloys when the three interfacial energies are equal ($\kappa_{AB} = \kappa_{BC} = \kappa_{AC}$), and so are the three species mobilities ($M_{A} = M_{B} = M_{C}$).

\subsection{Phase separation in an equimolar alloy: P1}
\label{sec_P1}
Figure~\ref{2a} and Figure~\ref{3a} show the microstructure evolution for Alloy P1, in which all three species have the same composition ($C_A=C_B=C_C=1/3$). Since the bulk free energy function is symmetric with respect to the three components, the volume fractions of the three phases in this alloy are also expected to be equal. One can clearly see the simultaneous formation of all three phases and their subsequent growth in both 2D and 3D, with roughly equal volume fractions of $\alpha, \beta$, and $\gamma$.

We have already emphasized the role of symmetric composition in Alloy P1. However, compositional asymmetry ($C_A \neq C_B = C_C$) can lead to a two-stage phase-separation process. We identify two types of such behavior. In Type I two-stage separation, the initial stage forms $B$-rich and $C$-rich regions, followed by the segregation of $A$ primarily at their interfaces. In Type II two-stage separation, the first stage produces $A$-rich and $A$-poor regions, with the second stage separation occurring primarily within the $A$-poor regions. Both cases are described below.


\subsection{Two stage spinodal decomposition: Type I}
\label{sec_A1}

Alloy A1 has a composition $(c_{A}, c_{B}, c_{C}) = (0.20, 0.40, 0.40)$. Since it is significantly poorer in $A$ compared to Alloy P1, the phase separation in A1 follows the Type I mechanism: initial formation of $B$-rich and $C$-rich regions (which evolve into the $\beta$ and $\gamma$ phases), followed by limited segregation of $A$ into interfacial regions, forming the $\alpha$ phase. As shown in Figure~\ref{2b} and Figure~\ref{3b}, A1 exhibits a lower volume fraction of the $\alpha$ phase compared to $\beta$ and $\gamma$, with $\alpha$ appearing as an interfacial phase embedded within a matrix formed by both $\beta$ and $\gamma$.

\subsection{Two stage spinodal decomposition: Type II}
\label{sec_A4}
Alloy A3 has a composition $(c_{A}, c_{B}, c_{C}) = (0.50, 0.25, 0.25)$. Since it is significantly richer in species $A$ than Alloy P1, the first stage of spinodal decomposition (SD) leads to the formation of $A$-rich and $A$-poor regions. The second stage is then triggered within the $A$-poor regions, where $B$-rich and $C$-rich domains emerge, eventually giving rise to the $\beta$ and $\gamma$ phases. As shown in Figure~\ref{2c} and Figure~\ref{3c}, A3 exhibits a higher volume fraction of the $\alpha$ phase (rich in $C_A$) compared to $\beta$ and $\gamma$. The resulting microstructure features an elongated, interconnected morphology of the $\alpha$ phase, with $\beta$ and $\gamma$ phases appearing as equiaxed islands embedded within it. These islands alternate along the elongated regions, suggesting a two-stage decomposition process within the $A$-poor matrix.

\section{Microstructure evolution in 2D and 3D: ML implementation}
\label{mlmethod}
\begin{figure}
\centering
\subfigure[$P1, t=1000$ composition map]{
\includegraphics[width=0.3\columnwidth]{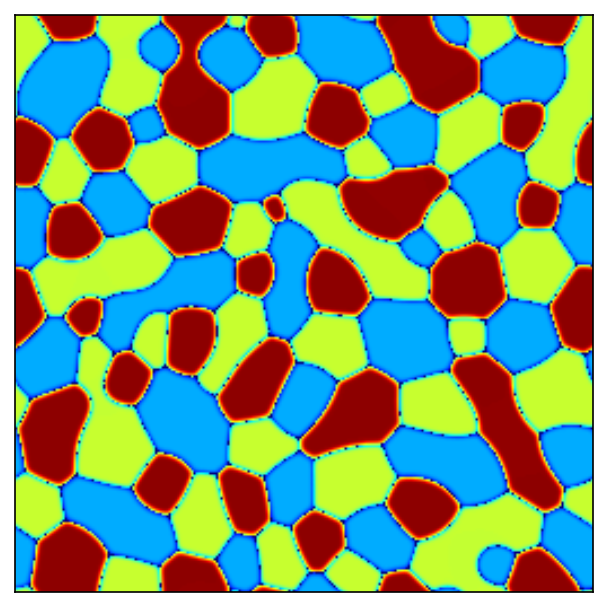}} 
\hfill
\subfigure[$P1, t=1000$ phase map]{
\includegraphics[width=0.3\columnwidth]{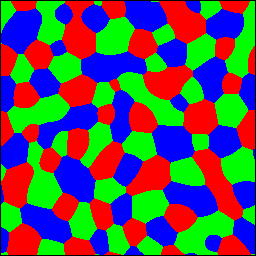}}\\
\subfigure[$A3, t=250$ composition map]{
\includegraphics[width=0.35\columnwidth]{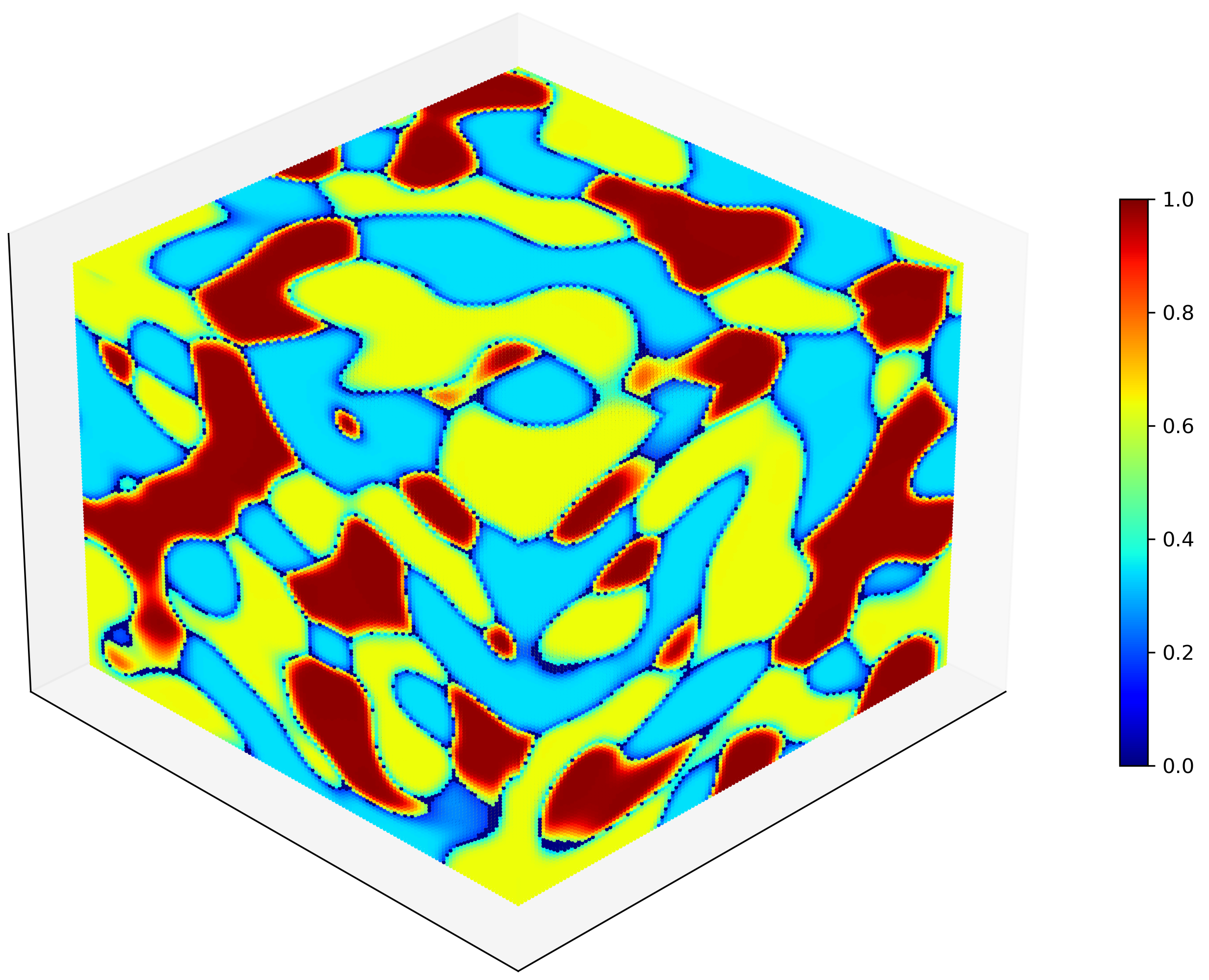}} 
\hfill
\subfigure[$A3, t=250$ phase map]{
\includegraphics[width=0.35\columnwidth]{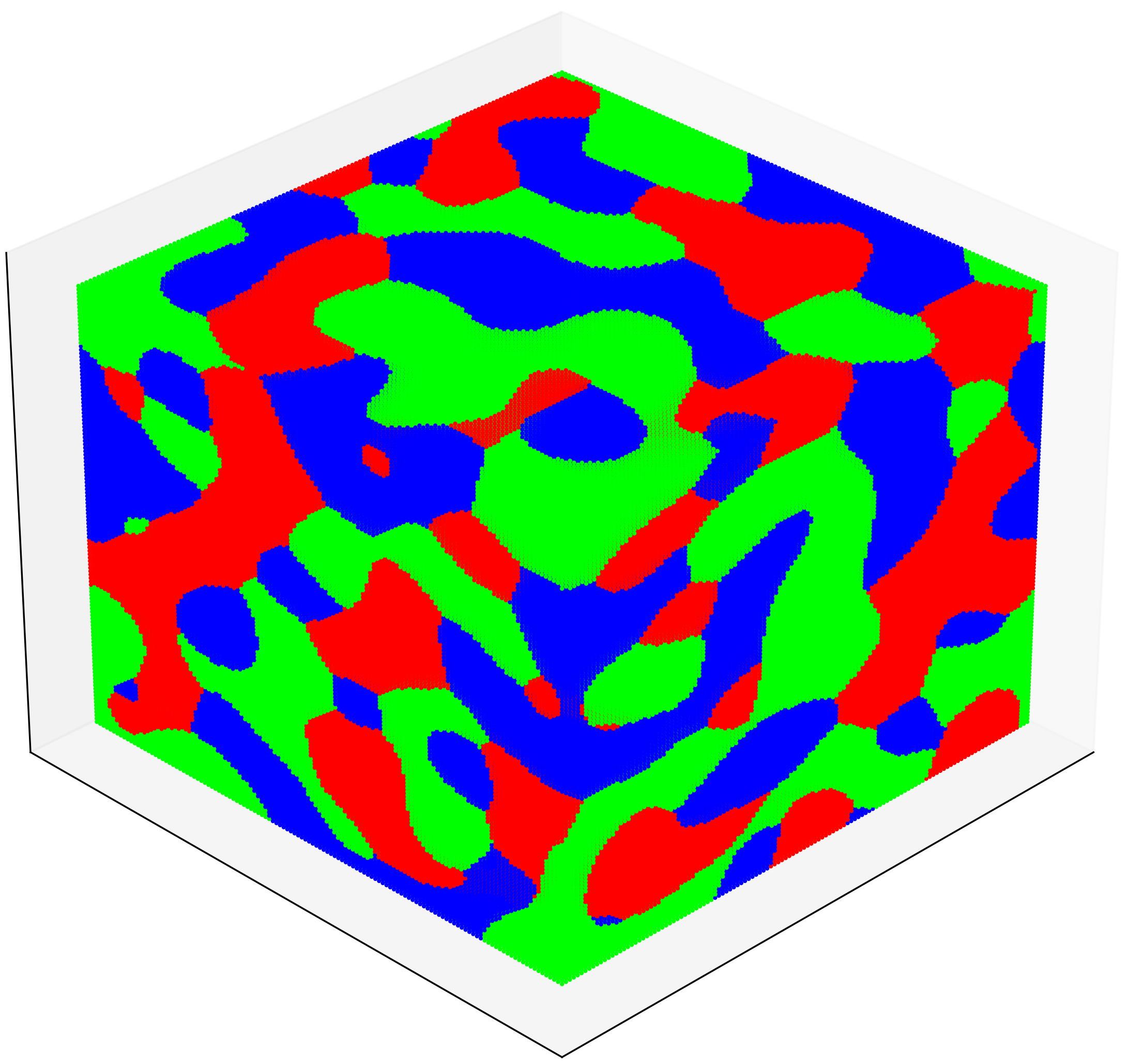}}
\caption{Composition and phase map comparison for both 2D and 3D.} 
\label{maps1}
\end{figure}

Having demonstrated the computational advantages of GPU-accelerated phase-field simulations, we now present our deep learning framework, which further accelerates predictions of microstructure evolution while maintaining accuracy. \textcolor{black}{While phase-field simulations by solving Cahn-Hilliard equations generate continuous compositional fields, our deep learning approach aims to infer discrete phase distributions based on these compositions. Specifically, we train our model to predict phase maps that indicate the locally dominant phase: A-rich (red), B-rich (green), or C-rich (blue), across the simulation domain, given a sequence of composition maps as input. This distinction is illustrated in Figure~\ref{maps1}, where the composition map represents the spatial distribution of component concentrations, and the corresponding phase map highlights the prevailing phase at each location. By learning this mapping from composition to phase, our framework enables rapid and accurate prediction of microstructural phase evolution across varying compositions.}

The phase-field method offers detailed temporal sequences of microstructural evolution, capturing complex transformation pathways from the initial to the final state. While Long Short-Term Memory (LSTM) networks have proven effective for sequence prediction tasks, the high dimensionality of microstructural image data ($256{\times}256{\times}3$ for 2D and $128{\times}128{\times}128{\times}3$ for 3D) poses significant challenges for efficient learning and prediction~\cite{Bostanabad2018}. We implement a three-part solution to address this challenge: dimensionality reduction through an autoencoder, attention-enhanced temporal prediction, and a novel slice-by-slice approach for three-dimensional phase predictions.

The core of our framework is an autoencoder architecture that efficiently compresses the high-dimensional microstructure data into a compact latent space representation while preserving essential physical features. The encoder function maps the input microstructure to this lower-dimensional space, while the decoder function reconstructs the original microstructure from the latent representation~\cite{autoencoderli}. This approach proves particularly effective for microstructure data, outperforming traditional dimensionality reduction techniques, such as Principal Component Analysis (PCA)~\cite{owaisprm}, in both compression efficiency and feature preservation.

The attention-enhanced ConvLSTM operates in the compressed latent space, where it learns to predict sequences of future microstructural states. The attention mechanism enables the model to focus on the most relevant spatial and temporal features for each prediction step, significantly improving accuracy, particularly for long-range predictions. The model takes five consecutive microstructures as input for each prediction sequence and generates predictions for subsequent states. These predictions, made in the latent space, are then decoded back to full-resolution microstructures.

Our framework extends naturally to three-dimensional predictions through a slice-by-slice approach, maintaining spatial coherence while leveraging the power of our 2D-trained model. The following sections detail the technical implementation of each component, including the autoencoder architecture, attention mechanism, and our approach to three-dimensional prediction.

\begin{table}
\centering
\caption{Encoder-Decoder architecture configuration. The input images have $256\times 256\times 3$ dimensions [$256\times 256$ pixels and three channels (RGB)].}
\footnotesize
\begin{tabular}{@{}llc@{}}
\hline
\noalign{\vskip 2pt} 
\textbf{Module} & \textbf{Layer Configuration} & \textbf{Output Shape} \\
\hline
\noalign{\vskip 2pt}
Encoder & ConvLSTM (32 filters, 3$\times$3 kernels) & (256,256,32) \\
Block 1 & LN + ReLU + Dropout(0.1) & (256,256,32) \\
& MaxPooling & (128,128,32) \\[2pt]
\hline
\noalign{\vskip 2pt}
Encoder & ConvLSTM (64 filters, 3$\times$3 kernels) & (128,128,64) \\
Block 2 & LN + Dropout (0.1) & (128,128,64) \\
& MaxPooling & (64,64,64) \\[2pt]
\hline
\noalign{\vskip 2pt}
Encoder & ConvLSTM (128 filters, 3$\times$3 kernels) & (64,64,128) \\
Block 3 & LN + Dropout (0.1) & (64,64,128) \\
& MaxPooling & (32,32,128) \\[2pt]
\hline
\noalign{\vskip 2pt}
Decoder & ConvLSTM (128 filters) & (32,32,128) \\
Block 1 & LN + Dropout (0.1) + Upsample & (64,64,128) \\[2pt]
\hline
\noalign{\vskip 2pt}
Decoder & ConvLSTM (64 filters) & (64,64,64) \\
Block 2 & LN + Dropout (0.1) + Upsample & (128,128,64) \\[2pt]
\hline
\noalign{\vskip 2pt}
Decoder & ConvLSTM (32 filters) & (128,128,32) \\
Block 3 & LN + Dropout (0.1) + Upsample & (256,256,32) \\
& Conv2D (3, 3$\times$3) + Sigmoid & (256,256,3) \\[2pt]
\hline
\end{tabular}
\label{tab:architecture}
\end{table}

\subsection{ConvLSTM architecture}
Our deep learning framework utilizes a sophisticated encoder-decoder architecture, enhanced with an attention mechanism, to predict microstructural evolution. The model is designed to process and predict spatiotemporal patterns of microstructural evolution, accounting for both spatial correlations and temporal dependencies. The architecture accepts input sequences of five consecutive microstructure frames at $256{\times}256{\times}3$ resolution and generates predictions for the subsequent five frames, effectively learning the underlying physics of phase transformation and structural evolution. The encoder pathway implements a hierarchical feature-extraction strategy via three cascaded ConvLSTM blocks. Each block is carefully designed to capture progressively more abstract representations of the microstructural features. The decoder pathway mirrors the encoder's structure but operates in reverse, progressively reconstructing the spatial dimensions while maintaining temporal coherence. This symmetric design facilitates effective information flow and the efficient utilization of features. Encoder-decoder architecture is summarized in Table~\ref{tab:architecture}. Further details are provided in Sections S4-S5, SI.

\subsection{Attention mechanism}
\label{attention}
A critical enhancement to our architecture is the integration of a Bahdanau attention mechanism between the encoder and decoder. This mechanism enables the model to dynamically focus on relevant spatiotemporal features during prediction, significantly improving the accuracy of long-term evolution predictions. The attention mechanism operates on the encoded features through the following mathematical framework:
\begin{equation}
    e_{ij} = V\tanh(W_1s_{i-1} + W_2h_j)
\end{equation}
where $s_{i-1}$ is the decoder's previous hidden state (query), $h_j$ represents the encoder's hidden states (keys), and $W_1$, $W_2$, and $V$ are learnable parameter matrices. The attention weights are computed through softmax normalization:
\begin{equation}
    \alpha_{ij} = \frac{\exp(e_{ij})}{\sum_{k=1}^{T}\exp(e_{ik})}.
\end{equation}
The context vector is then computed as:
\begin{equation}
    c_i = \sum_{j=1}^{T}\alpha_{ij}h_j.
\end{equation}
Further details are provided in Section S6, SI.

\subsection{Loss Function and Training Strategy}
The model employs a sophisticated training strategy optimized for microstructure evolution prediction:
\begin{equation}
    \mathcal{L}_{\text{total}} = \mathcal{L}_{\text{MSE}} + \lambda\mathcal{L}_{\text{temporal}}.
\end{equation}
In the above expression, the mean square error (MSE) is defined as,
\begin{equation}
    \mathcal{L}_{\text{MSE}} = \frac{1}{N}\sum_{i=1}^{N}(y_i - \hat{y}_i)^2,
\end{equation}
and the temporal coherence term ($\mathcal{L}_{\text{temporal}}$) is defined as,
\begin{equation}
    \mathcal{L}_{\text{temporal}} = \frac{1}{N-1}\sum_{i=1}^{N-1}\|(y_{i+1} - y_i) - (\hat{y}_{i+1} - \hat{y}_i)\|^2.
\end{equation}
The temporal coherence term, with a weight of $\lambda=0.1$, ensures smooth transitions between consecutive frames, which is crucial for maintaining physical consistency in the predicted microstructure evolution.

We use phase-field microstructures from 12 different compositions [see Figure~\ref{spinodal}] to train our model. For each composition, microstructures from $t=20$ to $t=1000$ time frames are added in the training dataset. Training is implemented using the Adam optimizer with an initial learning rate of $10^{-3}$ and gradient clipping at norm 1.0 to prevent gradient explosion. The learning process is monitored via a custom visualization callback that renders attention-weight distributions every 5 epochs, providing insights into the model's feature focus evolution during training.


\begin{figure}
\includegraphics[width=\columnwidth]{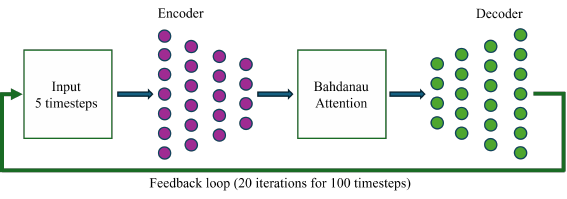}
\caption{Workflow of training a machine learning model with phase-field generated microstructures and building a machine-learned surrogate model for accelerated prediction of microstructure evolution. The model takes five consecutive microstructures as input for each prediction sequence and generates predictions for subsequent states. Starting with five phase-field microstructures and repeating the loop 20 times, one can predict the next 100 microstructures.}
\label{figml1}
\end{figure}

\begin{figure}
\includegraphics[width=\columnwidth]{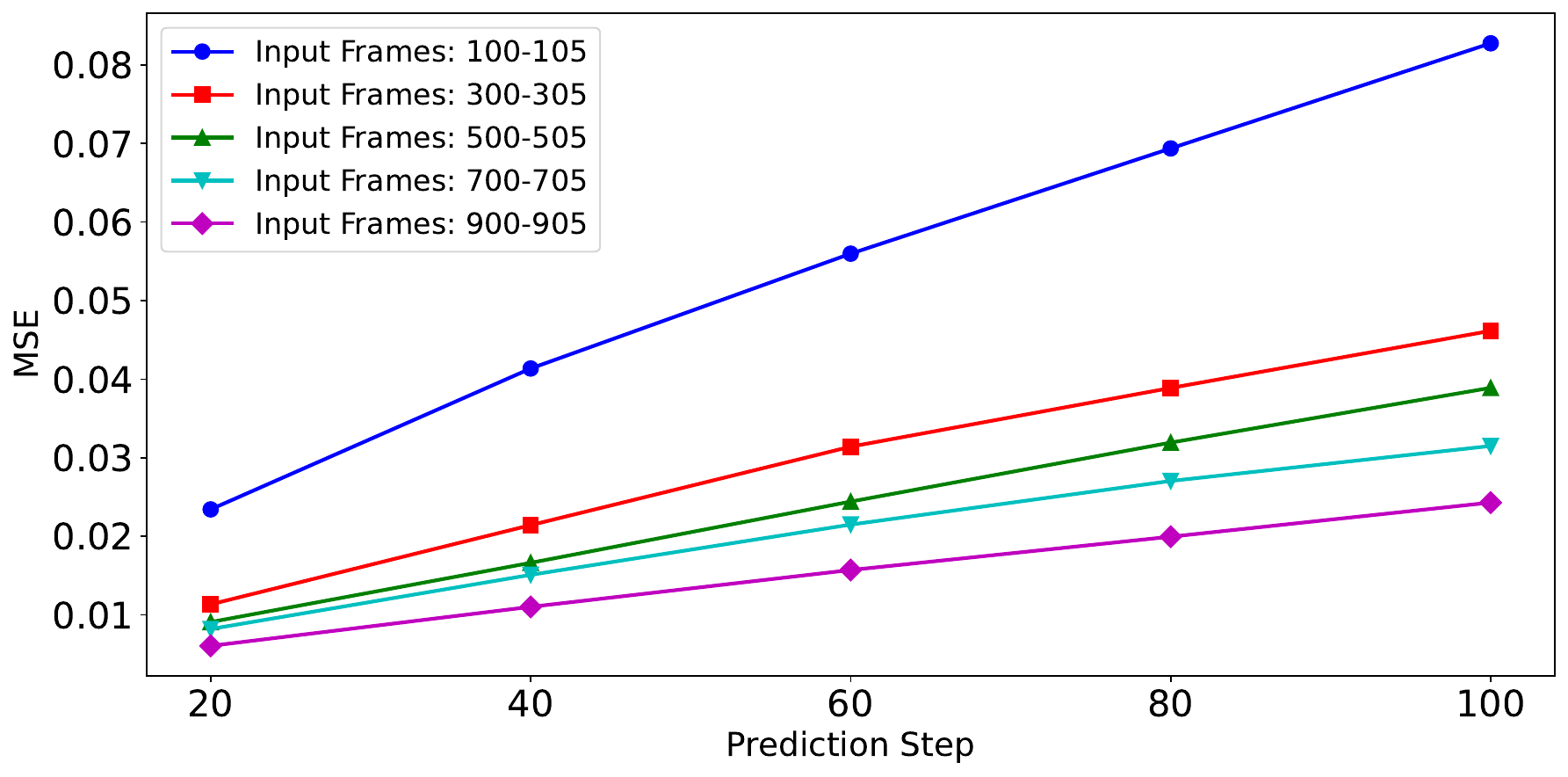}
\caption{Systematic analysis of prediction error accumulation across different stages of microstructure evolution. A systematic decrease in prediction errors is observed from early to late stages. This analysis is done for the composition A2 (0.40-0.30-0.30).}
\label{fig9}
\end{figure}


\subsection{Predicting 2D microstructures}
Among the predicted compositions [see Figure~\ref{spinodal} and Table~\ref{tabcomp} for the complete list], we present the results of A2 (0.40-0.30-0.30) in the main text, and the rest of the cases are illustrated in Figures~S5-S8, SI. As shown in Figure~\ref{figml1}, the model is provided with five phase-field microstructures, based on which it predicts the next 100 microstructures. We compare the predictive quality based on the MSE between the original (phase-field-generated) and the ML-predicted microstructure. We find that the quality of prediction improves at the later stage of microstructure evolution [Figure~\ref{fig9}]. For example, in the early stage (input time window $100-105$), the MSE values range from 0.025 ($20^{th}$ prediction) to 0.085 ($100^{th}$ prediction). On the other hand, at the late stage (input time window $900-905$), the MSE values range from 0.005 ($20^{th}$ prediction) to 0.020 ($100^{th}$ prediction). Clearly, the accuracy of the model improves at the later stages when the microstructural changes slow down. In the following paragraph, we investigate further by directly comparing the actual and predicted images.

\begin{figure*}
\includegraphics[width=0.48\linewidth]{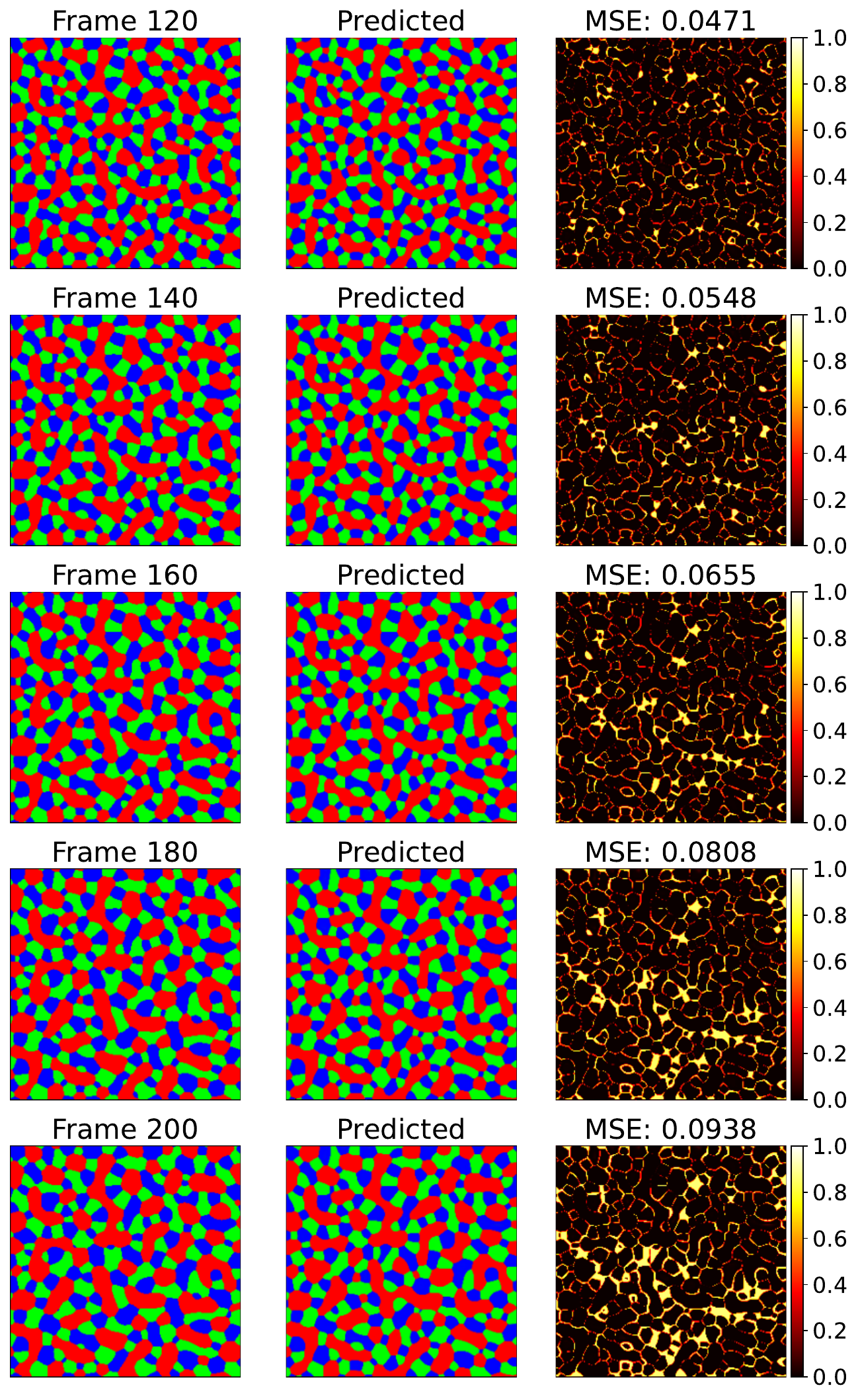}
\includegraphics[width=0.48\linewidth]{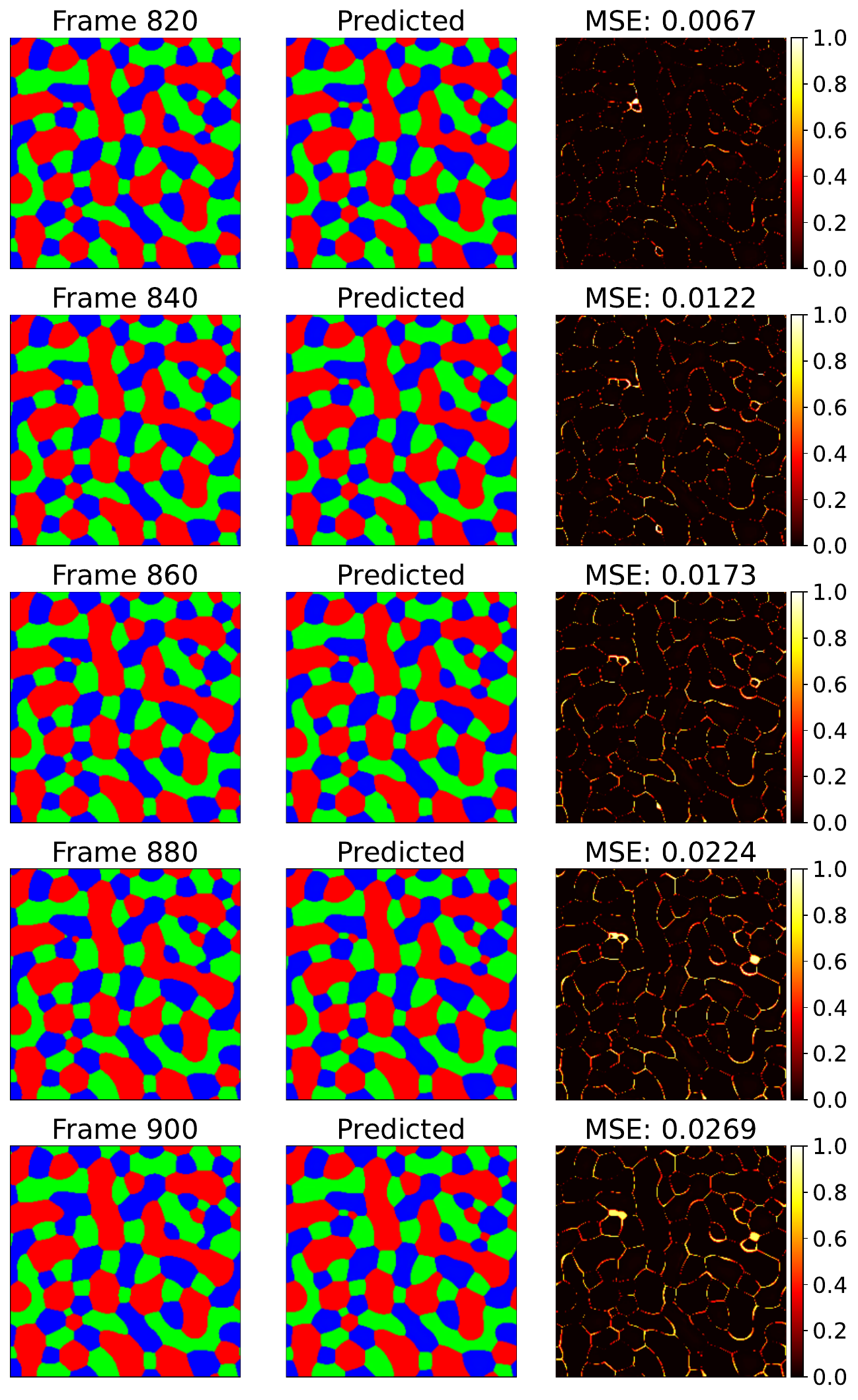}
\caption{Microstructure prediction for the composition A2 (0.40-0.30-0.30). (Left) Early-stage: Using five consecutive phase-field microstructures from $t=96$ to $t=100$ as input, the model predicts evolution up to $t=200$. (Right) Late-stage: Using five consecutive phase-field microstructures from $t=796$ to $t=800$ as input, the model predicts microstructure evolution up to $t=900$. The figures show original phase-field microstructures (first column), ML-predicted microstructures (middle column), and their difference maps (third column) at selected time steps. MSE indicates the deviation between the predicted and actual frames. Error compounding in ML-predicted microstructures is significantly smaller at later stages when the features change slowly over time.}
\label{fig7}
\end{figure*}

To evaluate the model's predictive accuracy at the early stage, we provide five consecutive phase-field microstructures from $t=96$ to $t=100$ as input, and the model predicts the next 100 frames of evolution. As shown in Figure~\ref{fig7} (left), the model successfully captures the microstructural changes. The mean squared error (MSE) between the predicted and actual microstructures ranges from 0.0514 (at $t=120$) to 0.0928 (at $t=200$), indicating reasonable accuracy.

For late-stage predictions, we provide five consecutive phase-field microstructures from $t=796$ to $t=800$ as input to predict the subsequent 100 frames. Figure~\ref{fig7} (right) demonstrates the model's significantly improved performance at this stage, with MSE values ranging from 0.0052 ($t=820$) to 0.0248 ($t=900$). This enhanced accuracy can be attributed to the slower, more gradual changes in microstructure during late-stage coarsening. We illustrate similar comparisons for other predicted compositions, as shown in Figures~S6-S9, SI.

\begin{figure*}
\includegraphics[width=0.46\linewidth]{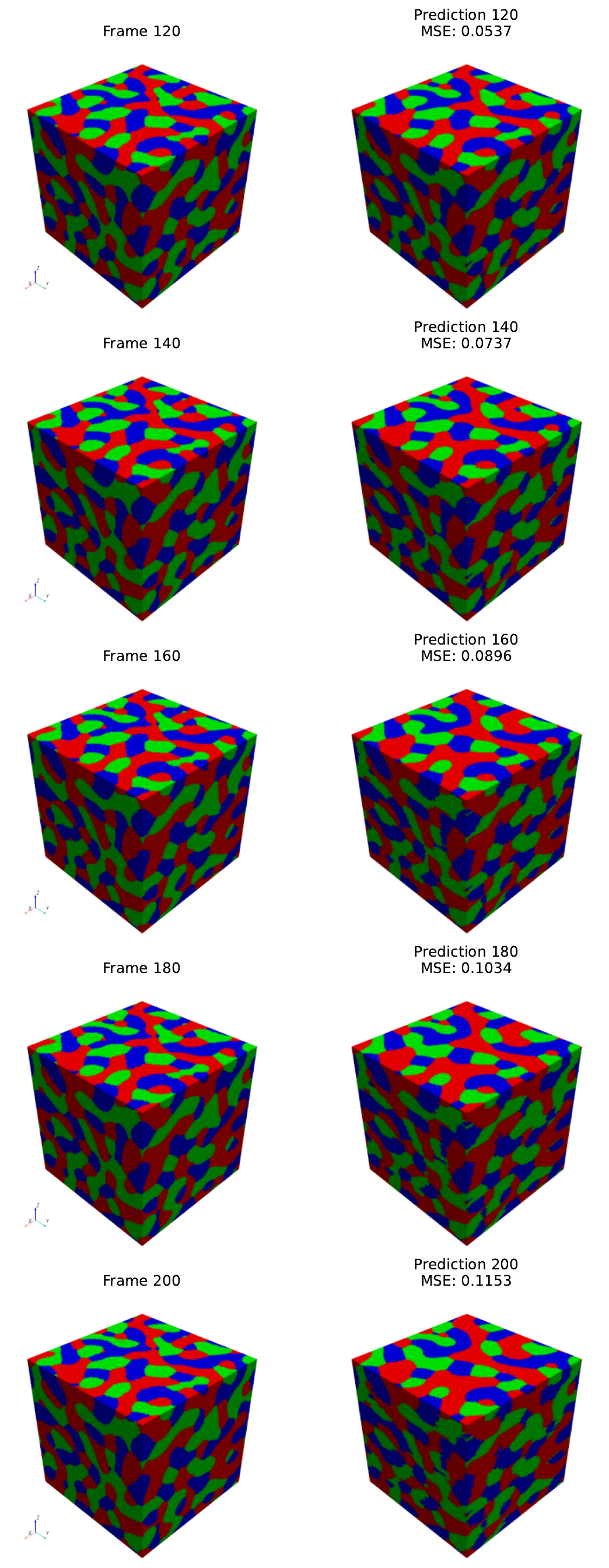}
\includegraphics[width=0.46\linewidth]{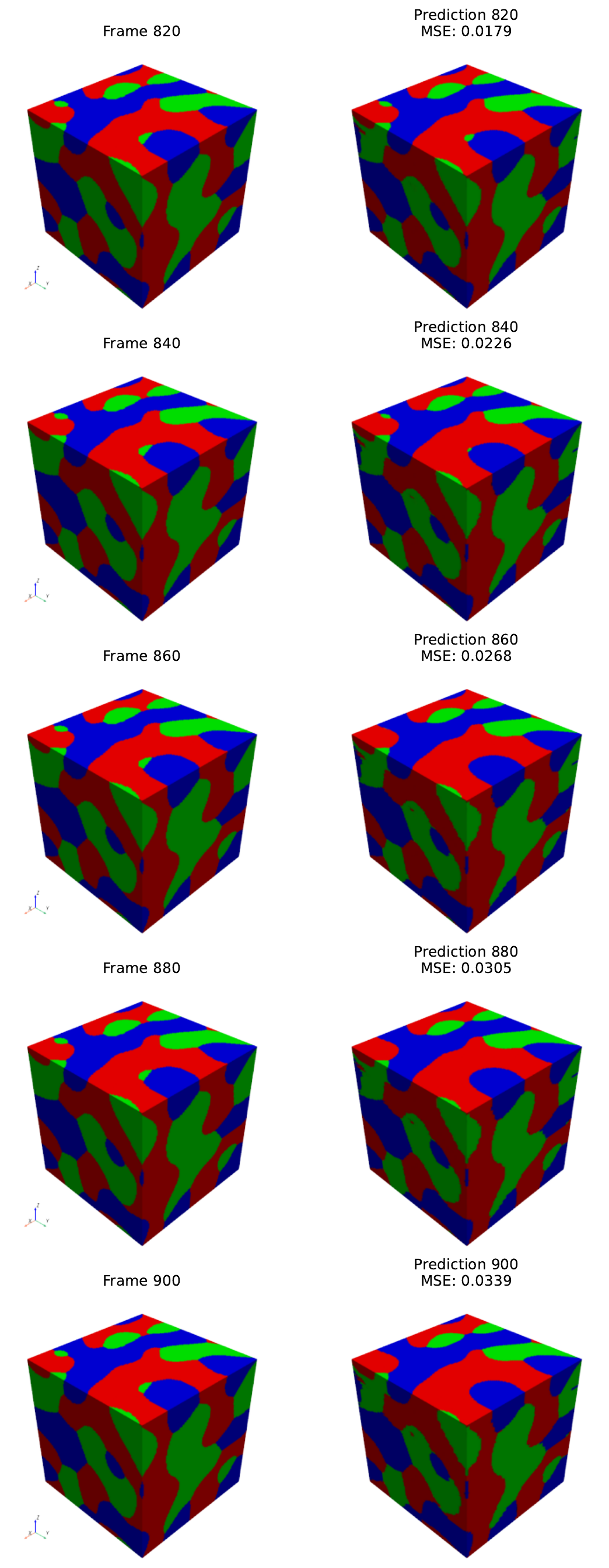}
\caption{Microstructure prediction for the composition A2 ($0.40-0.30-0.30$). (Left) Early-stage: 3D volume renderings of original phase-field simulations (first column) and predicted microstructures (second column) from $t=120$ to $t=200$, starting with five consecutive phase-field microstructures from $t=96$ to $t=100$. (Right) Late-stage: Predicted frames from $t=820$ to $t=900$, starting with five consecutive phase-field microstructures from $t=796$ to $t=800$. Similar to 2D, error compounding in ML-predicted microstructures is significantly smaller at the later stage.}
\label{fig10}
\end{figure*}

\subsection{Predicting 3D microstructures}
To extend our model's capabilities to three-dimensional microstructure evolution, we develop a slice-by-slice prediction approach that leverages our trained 2D model while maintaining spatial coherence across the third dimension. The input microstructure volume has dimensions of $128{\times}128{\times}128{\times}3$, where the first three dimensions represent the spatial coordinates and the last dimension contains the phase information.

Our slice-by-slice prediction strategy decomposes the 3D volume into a series of 2D slices along a single spatial axis. For each slice, we utilize the previously trained 2D model to generate predictions, maintaining temporal consistency through our attention-enhanced ConvLSTM architecture. The predicted 2D slices are then reassembled to reconstruct the complete 3D microstructure volume. This approach effectively balances computational efficiency with prediction accuracy by enabling us to leverage our well-trained 2D model while capturing the essential three-dimensional aspects of microstructure evolution.

To validate this approach, we conduct extensive testing on composition A2 (0.40-0.30-0.30) at two distinct evolutionary stages. Figure~\ref{fig10} (left) presents the early-stage predictions, where we use five consecutive 3D phase-field microstructures from $t=96$ to $t=100$ to predict the subsequent evolution up to $t=200$. The comparison between the predicted and actual phase-field microstructures demonstrates our model's ability to capture the complex three-dimensional evolution of microstructures. The mean squared error (MSE) between predicted and actual microstructures ranges from 0.0537 (at $t=120$) to 0.1153 (at $t=200$), which is marginally higher than that of 2D predictions.


Figure~\ref{fig10} (right) showcases the model's performance during late-stage evolution. Using 3D phase-field microstructures from $t=796$ to $t=800$, we predict evolution up to $t=900$. The mean squared error (MSE) between the predicted and actual microstructures improves remarkably, ranging from 0.0179 (at $t=820$) to 0.0339 (at $t=900$). We illustrate similar comparisons for other predicted compositions, as shown in Figures~S10-S13, SI. 

A notable observation from both early and late-stage predictions is that the slice-by-slice approach successfully maintains continuity across the third dimension, avoiding artificial discontinuities that might arise from treating each slice independently. This suggests that the spatial correlations learned by our 2D model effectively translate to three-dimensional microstructure evolution, particularly during the more stable late stages of phase separation. Similar to the case of our 2D model, 3D predictions are more accurate during the slow, late-stage coarsening dynamics.

\begin{figure}
\includegraphics[width=0.9\linewidth]{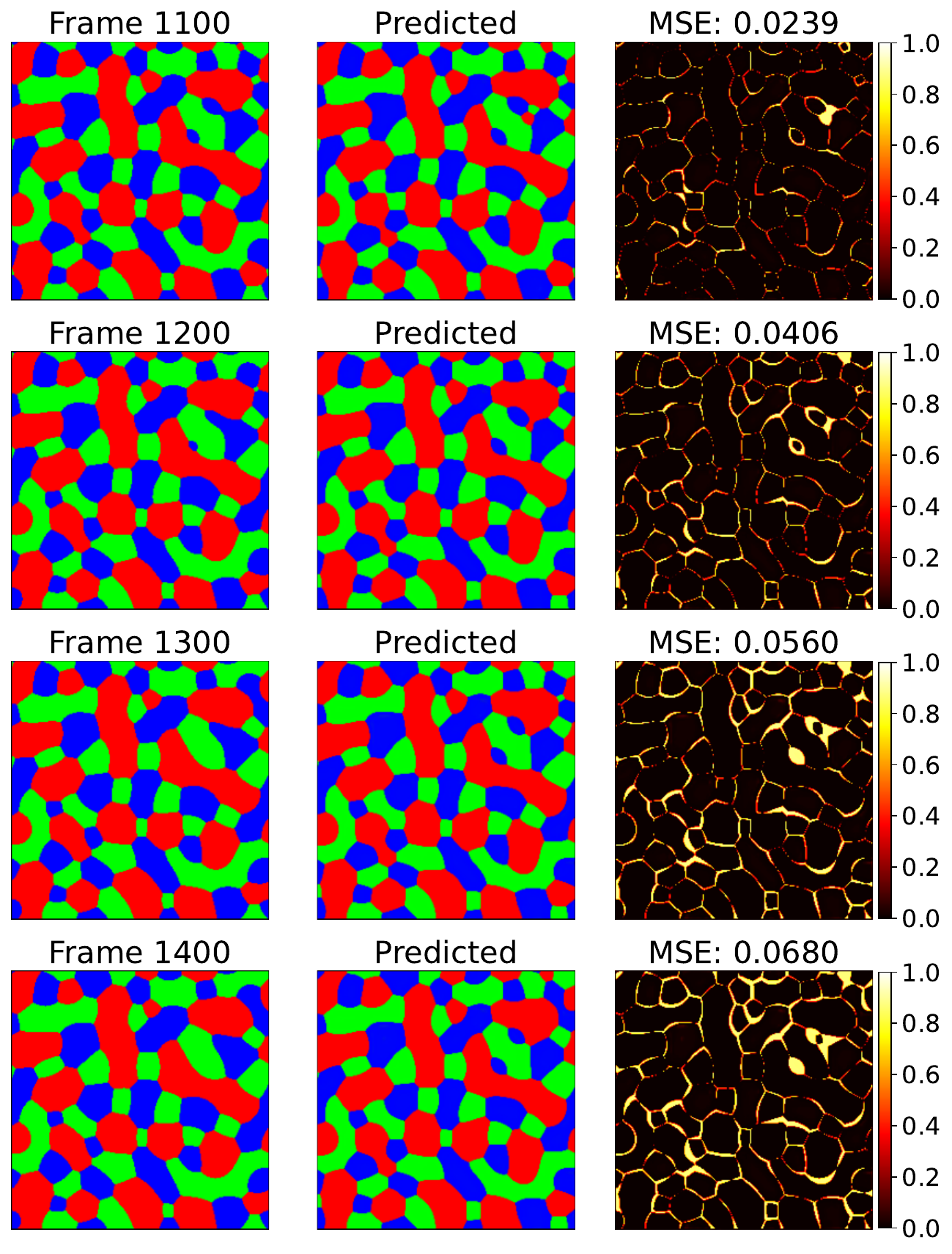}
\caption{Long-term 2D predictions for the composition A2 (0.40-0.30-0.30). Starting with input frames from $t=996$ to $t=1000$, the model predicts microstructure evolution up to $t=1400$, shown at 100-timestep intervals. The first column displays phase-field simulation results, the second column presents model predictions, and the third column shows difference maps.}
\label{fig12}
\end{figure}

\begin{figure}
\centering
\includegraphics[width=0.5\linewidth]{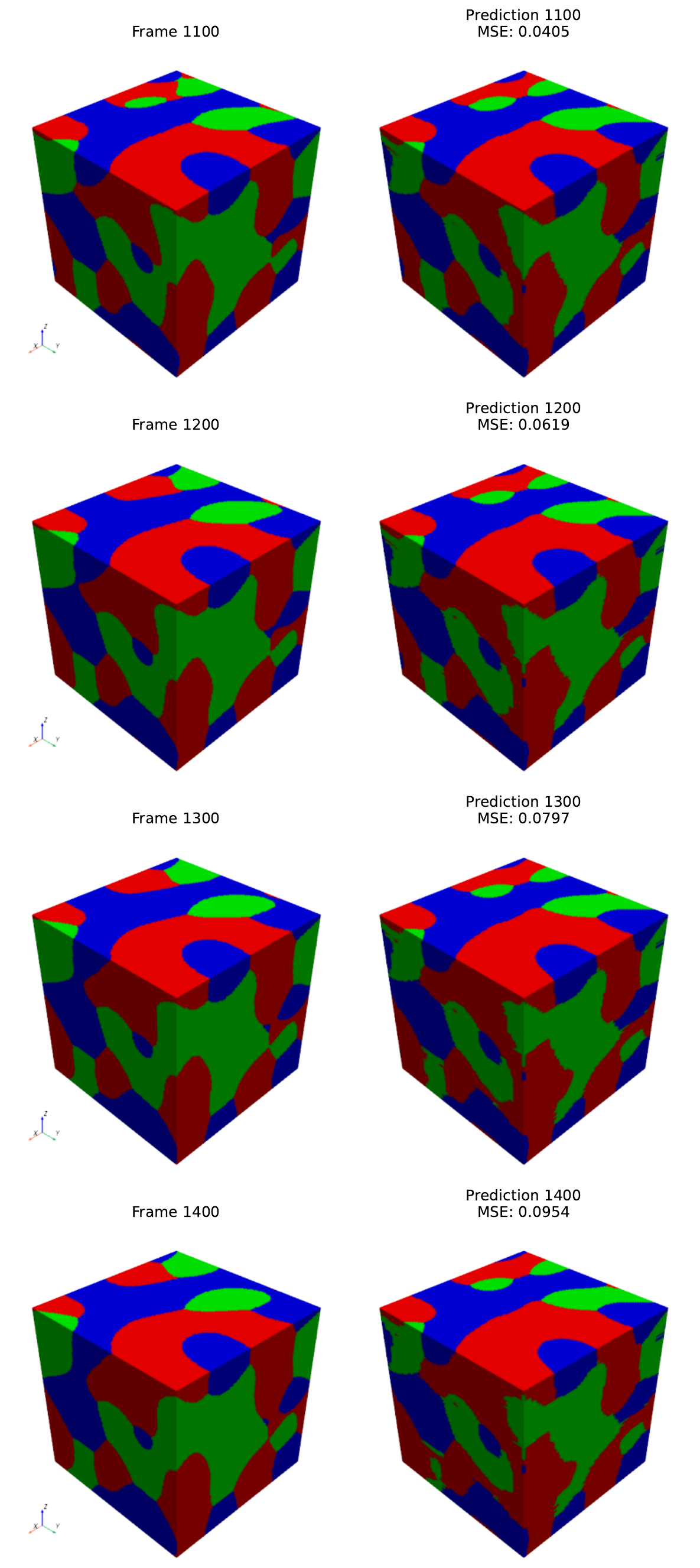}
\caption{Long-term 3D predictions for the composition A2 (0.40-0.30-0.30). Starting with input frames from $t=996$ to $t=1000$, the model predicts microstructure evolution up to $t=1400$, shown at 100-time-step intervals.}
\label{fig13}
\end{figure}

\subsection{Long-term predictions}
So far, we have successfully tested our model for predicting microstructures of any composition (not necessarily part of the training dataset) up to $t=1000$. In this section, we test the applicability of our model beyond $t=1000$. This is significant because our training dataset consists of microstructures up to the $1000^{th}$ time step. This exercise aims to test the generality of our model by extending the prediction horizon to forecast both unknown compositions and times beyond the training range.

We choose a composition A2 (0.40-0.30-0.30) and predict microstructure evolution up to $t=1400$, starting with five phase-field microstructures from $t=996$ to $t=1000$. Unlike the previous two sections (which were limited to predicting 100 future steps), we now attempt to forecast microstructures up to 400 time steps ahead. Figure~\ref{fig12} presents the predicted 2D microstructures, visualized at 100-time-step intervals to track the time evolution. Despite operating in a region of unseen composition and time, the model maintains remarkable accuracy. The MSE values range from 0.0239 ($t=1000$) to 0.0680 ($t=1400$). This relatively low error, even after 400 prediction steps, demonstrates the model's robust understanding of the underlying physics and its reliable long-term prediction capability, particularly during the coarsening process.

We perform a similar study for 3D predictions using a slice-by-slice approach. Figure~\ref{fig13} illustrates the predicted microstructures up to $t=1400$, using five phase-field microstructures ($t=996$ to $t=1000$) as input. The MSE values range from 0.0405 ($t=1100$) to 0.0954 ($t=1400$). The 3D predictions exhibit slightly higher error accumulation compared to the 2D predictions. The increased error compared to the 2D case can be attributed to the additional complexity of maintaining spatial coherence across the third dimension over such an extended prediction period. Nevertheless, the model successfully captures the major features of phase evolution, including the characteristic coarsening behavior and maintenance of phase volume fractions. These results highlight the model's ability to generalize to compositions and time outside its training set. 

\section{Conclusions}
\label{conclusion}
In this work, we demonstrate the effectiveness of a hybrid framework that combines phase-field simulation with AI/ML-assisted techniques to accelerate the modeling of 2D and 3D microstructure evolution in ternary three-phase alloys for nanotemplate application. The following key conclusions can be drawn from this study.

\begin{enumerate}
\item 
Three distinct phase-separation mechanisms occurring during the early stage of spinodal decomposition in ternary alloys are successfully captured in both 2D and 3D using composition maps.


\item 
The early-stage rapid dynamics are effectively captured using GPU-accelerated phase-field solutions, while an attention-enhanced ConvLSTM (Convolutional Long Short-Term Memory) model accurately predicts the late-stage coarsening behavior.

\item 
The performance advantage of the ML-model stems from two key architectural components: first, the autoencoder efficiently reduces the high-dimensional microstructure data ($256{\times}256{\times}3$ for 2D and $128{\times}128{\times}128{\times}3$ for 3D) to a compact latent representation while preserving essential phase transformation features. Second, the attention-enhanced ConvLSTM effectively learns and predicts spatiotemporal patterns of phase evolution, achieving remarkable accuracy, especially in capturing gradual changes during late-stage coarsening.

\item 
Our slice-by-slice approach for three-dimensional predictions represents a significant advancement, enabling the application of well-trained 2D models to 3D microstructure evolution while maintaining spatial coherence across all dimensions. The method's success, particularly evident in late-stage predictions with MSE as low as 0.02, suggests that the fundamental physics of phase separation is effectively captured by our model architecture. 

\item
The use of ML provides significant acceleration. Using an NVIDIA RTX A5000 GPU with 24 GB of device memory,
it takes the ML model $0.04$ seconds for predicting a 2D frame ($256 \times 256 \times 3$), and $14.03$ seconds for predicting a 3D frame ($128 \times 128 \times 128 \times 3$). Using similar resources, the GPU-accelerated phase-field model takes 0.456 seconds to generate a 2D frame and 29.9 seconds to generate a 3D frame.

\end{enumerate}

In summary, the demonstrated balance between computational efficiency and prediction accuracy positions this methodology as a valuable tool for studying and predicting microstructure evolution across a range of materials science and engineering fields such as producing nanoporous multiphase structures for catalysis applications.

\section{Acknowledgements}
The authors acknowledge the National Supercomputing Mission (NSM) for providing computing resources at Param Sanganak, IIT Kanpur, which is implemented by C-DAC and supported by the Ministry of Electronics and Information Technology (MeitY) and the Department of Science and Technology (DST), Government of India. RM and SB are thankful for the financial support received from the Center for Development of Advanced Computing (C-DAC) Project No. Meity/R\&D/HPC/2(1)/2014. The authors also acknowledge the ICME National Hub at IIT Kanpur for providing the computational facility.

\bibliography{ref2}

@article{doi:10.1021/acsami.5c13603,
author = {Chen, Xin and Yang, Lin and Zhang, Yuan and Zhang, Donglin and Song, Jiaqi and Feng, Jiaqi and Zhao, Xiuchen and Zhang, Gang and Huo, Yongjun},
title = {A Comprehensive Model for Spinodal Decomposition in Ag–Cu Alloys Based on Phase-Field Theory and In Situ TEM},
journal = {ACS Applied Materials \& Interfaces},
volume = {17},
number = {38},
pages = {54263-54281},
year = {2025},
doi = {10.1021/acsami.5c13603},
    note ={PMID: 40936195},
URL = {https://doi.org/10.1021/acsami.5c13603},
eprint = {https://doi.org/10.1021/acsami.5c13603}
}

@article{doi:10.1021/acsami.4c03011,
author = {Feng, Longsheng and Huang, Sijia and Heo, Tae Wook and Biener, Juergen},
title = {Integrated Framework to Model Microstructure Evolution and Decipher the Microstructure–Property Relationship in Polymeric Porous Materials},
journal = {ACS Applied Materials \& Interfaces},
volume = {16},
number = {29},
pages = {38442-38457},
year = {2024},
doi = {10.1021/acsami.4c03011},
    note ={PMID: 39009042},
URL = {https://doi.org/10.1021/acsami.4c03011},
eprint = {https://doi.org/10.1021/acsami.4c03011}
}

@Article{Ahmad_2025,
  author    = {Ahmad, Owais and Panwar, Vishal and Das, Kaushik and Mukherjee, Rajdip and Bhowmick, Somnath},
  journal   = {Physica Scripta},
  title     = {Scope of generative artificial intelligence in microstructural studies: a case study},
  year      = {2025},
  month     = {jul},
  number    = {7},
  pages     = {076020},
  volume    = {100},
  doi       = {10.1088/1402-4896/ade832},
  publisher = {IOP Publishing},
}

@Article{XU2020105303,
  author   = {Bo Xu and Guozheng Kang and Qianhua Kan and Chao Yu and Xi Xie},
  journal  = {International Journal of Mechanical Sciences},
  title    = {Phase field simulation on the cyclic degeneration of one-way shape memory effect of NiTi shape memory alloy single crystal},
  year     = {2020},
  issn     = {0020-7403},
  pages    = {105303},
  volume   = {168},
  doi      = {10.1016/j.ijmecsci.2019.105303},
  url      = {https://www.sciencedirect.com/science/article/pii/S0020740319330164},
}

@Article{ZHANG2019105008,
  author   = {Peng Zhang and Xiaofei Hu and Tinh Quoc Bui and Weian Yao},
  journal  = {International Journal of Mechanical Sciences},
  title    = {Phase field modeling of fracture in fiber reinforced composite laminate},
  year     = {2019},
  issn     = {0020-7403},
  pages    = {105008},
  volume   = {161-162},
  doi      = {10.1016/j.ijmecsci.2019.07.007},
  url      = {https://www.sciencedirect.com/science/article/pii/S0020740318341729},
}

@Article{LI2020105633,
  author   = {Pengfei Li and Julien Yvonnet and Christelle Combescure},
  journal  = {International Journal of Mechanical Sciences},
  title    = {An extension of the phase field method to model interactions between interfacial damage and brittle fracture in elastoplastic composites},
  year     = {2020},
  issn     = {0020-7403},
  pages    = {105633},
  volume   = {179},
  doi      = {10.1016/j.ijmecsci.2020.105633},
  url      = {https://www.sciencedirect.com/science/article/pii/S0020740320301211},
}

@Article{ZHOU2021106349,
  author   = {Shiwei Zhou and Yi Min Xie},
  journal  = {International Journal of Mechanical Sciences},
  title    = {Numerical simulation of three-dimensional multicomponent Cahn–Hilliard systems},
  year     = {2021},
  issn     = {0020-7403},
  pages    = {106349},
  volume   = {198},
  doi      = {10.1016/j.ijmecsci.2021.106349},
  url      = {https://www.sciencedirect.com/science/article/pii/S0020740321000849},
}

@Article{gaikwad2025,
  author    = {Gaikwad, Sachin and Kasilingam, Thejas and Ahmad, Owais and Mukherjee, Rajdip and Bhowmick, Somnath},
  journal   = {Phys. Rev. Mater.},
  title     = {Deep learning-driven prediction of microstructure evolution via latent space interpolation},
  year      = {2025},
  month     = {Oct},
  pages     = {103804},
  volume    = {9},
  doi       = {10.1103/5ngk-4v9j},
  issue     = {10},
  numpages  = {11},
  publisher = {American Physical Society},
}

@Article{TIWARI2025113518,
  author   = {Saurabh Tiwari and Prathamesh Satpute and Supriyo Ghosh},
  journal  = {{Computational Materials Science}},
  title    = {{Time Series Forecasting Of Multiphase Microstructure Evolution Using Deep Learning}},
  year     = {2025},
  issn     = {0927-0256},
  pages    = {113518},
  volume   = {247},
  doi      = {10.1016/j.commatsci.2024.113518},
  url      = {https://www.sciencedirect.com/science/article/pii/S0927025624007390},
}

@Article{YANG2017133,
  author   = {Cong Yang and Qingyan Xu and Baicheng Liu},
  journal  = {{Computational Materials Science}},
  title    = {{GPU-accelerated Three-dimensional Phase-field Simulation Of Dendrite Growth In A Nickel-based Superalloy}},
  year     = {2017},
  issn     = {0927-0256},
  pages    = {133-143},
  volume   = {136},
  doi      = {10.1016/j.commatsci.2017.04.031},
  url      = {https://www.sciencedirect.com/science/article/pii/S0927025617302239},
}

@Article{autoencoderli,
  author   = {Pengzhi Li and Yan Pei and Jianqiang Li},
  journal  = {{Applied Soft Computing}},
  title    = {{A Comprehensive Survey On Design And Application Of Autoencoder In Deep Learning}},
  year     = {2023},
  issn     = {1568-4946},
  pages    = {110176},
  volume   = {138},
  doi      = {10.1016/j.asoc.2023.110176},
  url      = {https://www.sciencedirect.com/science/article/pii/S1568494623001941},
}

@article{2002,
  title = {{Kinetics Of Phase Separation In Fe-Cr-Mo Ternary Alloys}},
  author={Yoshihiro Suwa and Yoshiyuki Saito and Kazumi Ochi and Takahiro Aoki and Kanako Goto and Kotaro Abe},
  journal = {{Materials Transactions}},
  volume={43},
  number={2},
  pages={271-276},
  year={2002},
  doi={10.2320/matertrans.43.271}
}

@Article{CHEN1998147,
  author   = {L.Q. Chen and Jie Shen},
  journal  = {{Computer Physics Communications}},
  title    = {{Applications Of Semi-implicit Fourier-spectral Method To Phase Field Equations}},
  year     = {1998},
  issn     = {0010-4655},
  number   = {2},
  pages    = {147-158},
  volume   = {108},
  doi      = {10.1016/S0010-4655(97)00115-X},
  url      = {https://www.sciencedirect.com/science/article/pii/S001046559700115X},
}

@Article{Park2012,
  author   = {Park, Jin Man and Han, Jun Hee and Mattern, Norbert and Kim, Do Hyang and Eckert, Jurgen},
  journal  = {{Metallurgical And Materials Transactions A}},
  title    = {{{Designing Zr-Cu-Co-Al Bulk Metallic Glasses With Phase Separation Mediated Plasticity}}},
  year     = {2012},
  issn     = {1543-1940},
  number   = {8},
  pages    = {2598-2603},
  volume   = {43},
  day      = {01},
  doi      = {10.1007/s11661-011-1050-z},
}

@Article{SOHN201257,
  author   = {S.W. Sohn and W. Yook and W.T. Kim and D.H. Kim},
  journal  = {{Intermetallics}},
  title    = {{{Phase Separation In Bulk-type Gd-Zr-Al-Ni Metallic Glass}}},
  year     = {2012},
  issn     = {0966-9795},
  pages    = {57-62},
  volume   = {23},
  doi      = {10.1016/j.intermet.2011.12.021},
  url      = {https://www.sciencedirect.com/science/article/pii/S0966979511004092},
}

@Article{CHANG20102483,
  author   = {H.J. Chang and W. Yook and E.S. Park and J.S. Kyeong and D.H. Kim},
  journal  = {{Acta Materialia}},
  title    = {{Synthesis Of Metallic Glass Composites Using Phase Separation Phenomena}},
  year     = {2010},
  issn     = {1359-6454},
  number   = {7},
  pages    = {2483-2491},
  volume   = {58},
  doi      = {10.1016/j.actamat.2009.12.034},
  url      = {https://www.sciencedirect.com/science/article/pii/S1359645409008829},
}

@Article{1Pattanayak,
  author   = {Pattanayak, Pulok and Asokan, S.},
  journal  = {{Journal Of Applied Physics}},
  title    = {{{Signature of A Silver Phase Percolation Threshold in Microscopically Phase Separated Ternary Ge$_{0.15}$Se$_{0.85-x}$Ag$_x$ $(0\leq x \leq 0.20)$ Glasses}}},
  year     = {2004},
  issn     = {0021-8979},
  number   = {1},
  pages    = {013515},
  volume   = {97},
  doi      = {10.1063/1.1827341},
}

@Article{MATTERN2010299,
  author   = {Norbert Mattern and Thomas Gemming and J{\"u}rgen Thomas and G{\"u}nter Goerigk and Hermann Franz and J{\"u}rgen Eckert},
  journal  = {{Journal Of Alloys And Compounds}},
  title    = {{{Phase Separation In Ni-Nb-Y Metallic Glasses}}},
  year     = {2010},
  issn     = {0925-8388},
  number   = {2},
  pages    = {299-304},
  volume   = {495},
  doi      = {10.1016/j.jallcom.2009.10.013},
  url      = {https://www.sciencedirect.com/science/article/pii/S0925838809019707},
}

@Article{MATTERN2009903,
  author   = {N. Mattern and G. Goerigk and U. Vainio and M.K. Miller and T. Gemming and J. Eckert},
  journal  = {{Acta Materialia}},
  title    = {{Spinodal Decomposition Of Ni-Nb-Y Metallic Glasses}},
  year     = {2009},
  issn     = {1359-6454},
  number   = {3},
  pages    = {903-908},
  volume   = {57},
  doi      = {10.1016/j.actamat.2008.10.028},
  url      = {https://www.sciencedirect.com/science/article/pii/S1359645408007787},
}

@Article{GOERIGK20093652,
  author   = {G. Goerigk and N. Mattern},
  journal  = {{Acta Materialia}},
  title    = {{Critical Scattering Of Ni-Nb-Y Metallic Glasses Probed By Quantitative Anomalous Small-angle X-ray Scattering}},
  year     = {2009},
  issn     = {1359-6454},
  number   = {12},
  pages    = {3652-3661},
  volume   = {57},
  doi      = {10.1016/j.actamat.2009.04.028},
  url      = {https://www.sciencedirect.com/science/article/pii/S135964540900250X},
}

@ARTICLE{okada1978,
  author={Okada, M. and Thomas, G. and Homma, M. and Kaneko, H.},
  journal = {{IEEE Transactions On Magnetics}}, 
  title = {{Microstructure And Magnetic Properties Of Fe-Cr-Co Alloys}}, 
  year={1978},
  volume={14},
  number={4},
  pages={245-252},
  doi={10.1109/TMAG.1978.1059752}}

@Article{chu1985,
  author   = {Chu, S. N. G. and Nakahara, S. and Strege, K. E. and Johnston, W. D., Jr.},
  journal  = {{Journal Of Applied Physics}},
  title    = {{{Surface Layer Spinodal Decomposition In In$_{1-x}$Ga$_x$As$_y$P$_{1-y}$ And In$_{1-x}$Ga$_x$ As Grown By Hydride Transport Vapor-phase Epitaxy}}},
  year     = {1985},
  issn     = {0021-8979},
  number   = {10},
  pages    = {4610--4615},
  volume   = {57},
  doi      = {10.1063/1.335368},
  eprint   = {https://pubs.aip.org/aip/jap/article-pdf/57/10/4610/18411496/4610\_1\_online.pdf},
}

@Article{LouiseMakin1984,
  author   = {Louise Makin, P. and Ralph, Brian},
  journal  = {{Journal Of Materials Science}},
  title    = {{On The Ageing Of An Aluminium-Lithium-Zirconium Alloy}},
  year     = {1984},
  issn     = {1573-4803},
  number   = {12},
  pages    = {3835-3843},
  volume   = {19},
  day      = {01},
  doi      = {10.1007/BF00980745},
}

@ARTICLE{spooner1980,
       author = {{Spooner}, S. and {Lefevre}, B.~G.},
        title = "{The effect of prior deformation on spinodal age hardening in Cu-15 Ni-8 Sn alloy}",
      journal = {{Metallurgical Transactions A}},
         year = 1980,
        
       volume = {11},
       number = {7},
        pages = {1085-1093},
          doi = {10.1007/BF02668132},
       adsurl = {https://ui.adsabs.harvard.edu/abs/1980MTA....11.1085S}
}

@Article{Singh1980,
  author   = {Singh, J. and Lele, S. and Ranganathan, S.},
  journal  = {{Journal Of Materials Science}},
  title    = {{Spinodal Decomposition In Co-3wt{\%} Ti-1wt{\%} Fe And Co-3 Wt{\%} Ti-2 Wt{\%} Fe Alloys}},
  year     = {1980},
  issn     = {1573-4803},
  number   = {8},
  pages    = {2010-2016},
  volume   = {15},
  day      = {01},
  doi      = {10.1007/BF00550627},
}

@Article{YIN2025110382,
  author   = {Yaode Yin and Hongjun Yu},
  journal  = {International Journal of Mechanical Sciences},
  title    = {A thermodynamic-consistent phase-field model for fracture in temperature-dependent materials},
  year     = {2025},
  issn     = {0020-7403},
  pages    = {110382},
  volume   = {297-298},
  doi      = {10.1016/j.ijmecsci.2025.110382},
  url      = {https://www.sciencedirect.com/science/article/pii/S0020740325004680},
}

@Article{Chen199415752,
  author        = {Chen, L.-Q. and Yang, W.},
  journal       = {Physical Review B},
  title         = {Computer simulation of the domain dynamics of a quenched system with a large number of nonconserved order parameters: The grain-growth kinetics},
  year          = {1994},
  number        = {21},
  pages         = {15752-15756},
  volume        = {50},
  document_type = {Article},
  doi           = {10.1103/physrevb.50.15752},
  source        = {Scopus},
}

@Article{Cahn1961795,
  author        = {Cahn, J. W.},
  journal       = {Acta Metallurgica},
  title         = {On spinodal decomposition},
  year          = {1961},
  number        = {9},
  pages         = {795-801},
  volume        = {9},
  document_type = {Article},
  doi           = {10.1016/0001-6160(61)90182-1},
  source        = {Scopus},
}

@Article{Cahn1962,
  author    = {John W. Cahn},
  journal   = {Acta Metallurgica},
  title     = {On spinodal decomposition in cubic crystals},
  year      = {1962},
  issn      = {0001-6160},
  month     = {3},
  pages     = {179-183},
  volume    = {10},
  abcd      = {10.1016/0001-6160(62)90114-1},
  doi       = {10.1016/0001-6160(62)90114-1},
  issue     = {3},
  publisher = {Pergamon},
}

@Article{Wang2004,
  author    = {Jie Wang and San Qiang Shi and Long Qing Chen and Yulan Li and Tong Yi Zhang},
  journal   = {Acta Materialia},
  title     = {Phase-field simulations of ferroelectric/ferroelastic polarization switching},
  year      = {2004},
  issn      = {1359-6454},
  month     = {2},
  pages     = {749-764},
  volume    = {52},
  abcd      = {10.1016/j.actamat.2003.10.011},
  doi       = {10.1016/j.actamat.2003.10.011},
  issue     = {3},
  publisher = {Pergamon},
}

@Article{Boettinger2002163,
  author        = {Boettinger, W.J. and Warren, J.A. and Beckermann, C. and Karma, A.},
  journal       = {Annual Review of Materials Science},
  title         = {Phase-field simulation of solidification},
  year          = {2002},
  pages         = {163-194},
  volume        = {32},
  document_type = {Review},
  doi           = {10.1146/annurev.matsci.32.101901.155803},
  source        = {Scopus},
}

@Article{PhysRevE.57.4323,
  author    = {Karma, Alain and Rappel, Wouter-Jan},
  journal   = {Phys. Rev. E},
  title     = {Quantitative phase-field modeling of dendritic growth in two and three dimensions},
  year      = {1998},
  month     = {Apr},
  pages     = {4323--4349},
  volume    = {57},
  abc       = {https://link.aps.org/abcd/10.1103/PhysRevE.57.4323},
  abcd      = {10.1103/PhysRevE.57.4323},
  doi       = {10.1103/physreve.57.4323},
  issue     = {4},
  numpages  = {0},
  publisher = {American Physical Society},
}

@Article{Chen2015,
  author    = {L. Chen and J. Chen and R. A. Lebensohn and Y. Z. Ji and T. W. Heo and S. Bhattacharyya and K. Chang and S. Mathaudhu and Z. K. Liu and L. Q. Chen},
  journal   = {Computer Methods in Applied Mechanics and Engineering},
  title     = {An integrated fast Fourier transform-based phase-field and crystal plasticity approach to model recrystallization of three dimensional polycrystals},
  year      = {2015},
  issn      = {0045-7825},
  month     = {3},
  pages     = {829-848},
  volume    = {285},
  abcd      = {10.1016/j.cma.2014.12.007},
  doi       = {10.1016/j.cma.2014.12.007},
  publisher = {Elsevier},
}

@Article{Fan2002,
  author    = {Danan Fan and S. P. Chen and Long Qing Chen and P. W. Voorhees},
  journal   = {Acta Materialia},
  title     = {Phase-field simulation of 2-D Ostwald ripening in the high volume fraction regime},
  year      = {2002},
  issn      = {1359-6454},
  month     = {5},
  pages     = {1895-1907},
  volume    = {50},
  abcd      = {10.1016/S1359-6454(01)00393-7},
  doi       = {10.1016/s1359-6454(01)00393-7},
  issue     = {8},
  publisher = {Pergamon},
}

@Article{Rodney2003,
  author    = {D. Rodney and Y. Le Bouar and Alphonse Finel},
  journal   = {Acta Materialia},
  title     = {Phase field methods and dislocations},
  year      = {2003},
  issn      = {1359-6454},
  month     = {1},
  pages     = {17-30},
  volume    = {51},
  abcd      = {10.1016/S1359-6454(01)00379-2},
  doi       = {10.1016/s1359-6454(01)00379-2},
  issue     = {1},
  publisher = {Pergamon},
}

@Article{Jin2001,
  author    = {Y. M. Jin and A. G. Khachaturyan},
  journal   = {Philosophical Magazine Letters},
  title     = {Phase field microelasticity theory of dislocation dynamics in a polycrystal: Model and three-dimensional simulations},
  year      = {2001},
  issn      = {0950-0839},
  month     = {9},
  pages     = {607-616},
  volume    = {81},
  abcd      = {10.1080/09500830110062825},
  doi       = {10.1080/09500830110062825},
  issue     = {9},
  publisher = {Taylor & Francis Group},
}

@Article{Henry2004,
  author    = {Hervé Henry and Herbert Levine},
  journal   = {Physical Review Letters},
  title     = {Dynamic instabilities of fracture under biaxial strain using a phase field model},
  year      = {2004},
  issn      = {0031-9007},
  month     = {9},
  pages     = {105504},
  volume    = {93},
  abcd      = {10.1103/PhysRevLett.93.105504},
  doi       = {10.1103/physrevlett.93.105504},
  issue     = {10},
  publisher = {American Physical Society},
}

@Article{Spatschek2011,
  author    = {Robert Spatschek and Efim Brener and Alain Karma},
  journal   = {Philosophical Magazine},
  title     = {Phase field modeling of crack propagation},
  year      = {2011},
  issn      = {1478-6435},
  month     = {1},
  pages     = {75-95},
  volume    = {91},
  abc       = {http://www.tandfonline.com/abcd/abs/10.1080/14786431003773015},
  abcd      = {10.1080/14786431003773015},
  doi       = {10.1080/14786431003773015},
  issue     = {1},
  publisher = {Taylor & Francis Group},
}

@Article{Leo1998,
  author    = {P. H. Leo and J. S. Lowengrub and H. J. Jou},
  journal   = {Acta Materialia},
  title     = {A diffuse interface model for microstructural evolution in elastically stressed solids},
  year      = {1998},
  issn      = {1359-6454},
  month     = {3},
  pages     = {2113-2130},
  volume    = {46},
  abcd      = {10.1016/S1359-6454(97)00377-7},
  doi       = {10.1016/s1359-6454(97)00377-7},
  issue     = {6},
  publisher = {Elsevier Ltd},
}

@Article{Gururajan2007,
  author   = {M. P. Gururajan and T. A. Abinandanan},
  journal  = {Philosophical Magazine},
  title    = {Phase inversion in two-phase solid systems driven by an elastic modulus mismatch},
  year     = {2007},
  issn     = {1478-6435},
  month    = {11},
  pages    = {5279-5288},
  volume   = {87},
  abc      = {http://www.tandfonline.com/abcd/abs/10.1080/14786430701647984},
  abcd     = {10.1080/14786430701647984},
  doi      = {10.1080/14786430701647984},
  issue    = {33},
}

@Article{Koyama2008,
  author    = {Toshiyuki Koyama},
  journal   = {Science and Technology of Advanced Materials},
  title     = {Phase-field modeling of microstructure evolutions in magnetic materials},
  year      = {2008},
  issn      = {1468-6996},
  month     = {1},
  pages     = {013006},
  volume    = {9},
  abc       = {https://www.tandfonline.com/abcd/full/10.1088/1468-6996/9/1/013006},
  abcd      = {10.1088/1468-6996/9/1/013006},
  doi       = {10.1088/1468-6996/9/1/013006},
  issue     = {1},
  publisher = {Taylor & Francis},
}

@Article{Mukherjee2016,
  author     = {Mukherjee, A. and Mukherjee, R. and Ankit, K. and Bhattacharya, A. and Nestler, B.},
  journal    = {Physical Review E},
  title      = {Influence of substrate interaction and confinement on electric-field-induced transition in symmetric block-copolymer thin films},
  year       = {2016},
  number     = {3},
  volume     = {93},
  art_number = {032504},
  doi        = {10.1103/physreve.93.032504},
}

@Article{Mukherjee20162,
  author    = {Arnab Mukherjee and Kumar Ankit and Andreas Reiter and Michael Selzer and Britta Nestler},
  journal   = {Physical Chemistry Chemical Physics},
  title     = {Electric-field-induced lamellar to hexagonally perforated lamellar transition in diblock copolymer thin films: Kinetic pathways},
  year      = {2016},
  issn      = {1463-9076},
  month     = {9},
  pages     = {25609-25620},
  volume    = {18},
  abcd      = {10.1039/c6cp04903f},
  doi       = {10.1039/c6cp04903f},
  issue     = {36},
  publisher = {Royal Society of Chemistry},
}

@Article{WU2025109792,
  author   = {Xin-Wei Wu and Mingyang Chen and Liao-Liang Ke},
  journal  = {International Journal of Mechanical Sciences},
  title    = {An electro-thermo-mechanical coupling phase-field model of defect evolution induced by electromigration in interconnects},
  year     = {2025},
  issn     = {0020-7403},
  pages    = {109792},
  volume   = {285},
  doi      = {10.1016/j.ijmecsci.2024.109792},
  url      = {https://www.sciencedirect.com/science/article/pii/S0020740324008336},
}

@Article{LIVAK1974589,
  author   = {R.J Livak and G Thomas},
  journal  = {{Acta Metallurgica}},
  title    = {{Loss Of Coherency In Spinodally Decomposed Cu-Ni-Fe Alloys}},
  year     = {1974},
  issn     = {0001-6160},
  number   = {5},
  pages    = {589-599},
  volume   = {22},
  doi      = {10.1016/0001-6160(74)90156-4},
  url      = {https://www.sciencedirect.com/science/article/pii/0001616074901564},
}

@Article{COPETTI200041,
  author   = {M.I.M. Copetti},
  journal  = {{Mathematics And Computers In Simulation}},
  title    = {{Numerical Experiments Of Phase Separation In Ternary Mixtures}},
  year     = {2000},
  issn     = {0378-4754},
  number   = {1},
  pages    = {41-51},
  volume   = {52},
  doi      = {10.1016/S0378-4754(99)00153-6},
  url      = {https://www.sciencedirect.com/science/article/pii/S0378475499001536},
}

@Article{qiang,
  author   = {Ma, Yu-qiang},
  journal  = {{The Journal Of Chemical Physics}},
  title    = {{Domain patterns in ternary mixtures with different interfacial properties}},
  year     = {2001},
  issn     = {0021-9606},
  number   = {8},
  pages    = {3734-3738},
  volume   = {114},
  doi      = {10.1063/1.1343838},
  eprint   = {https://pubs.aip.org/aip/jcp/article-pdf/114/8/3734/19016275/3734\_1\_online.pdf},
}

@Article{travasso,
  author   = {Travasso, Rui D. M. and Buxton, Gavin A. and Kuksenok, Olga and Good, Kevin and Balazs, Anna C.},
  journal  = {{The Journal Of Chemical Physics}},
  title    = {{Modeling the morphology and mechanical properties of sheared ternary mixtures}},
  year     = {2005},
  issn     = {0021-9606},
  number   = {19},
  pages    = {194906},
  volume   = {122},
  doi      = {10.1063/1.1903883},
}

@Article{alfarraj,
  author   = {Alfarraj, Abdulrahman A. and Nauman, E. Bruce},
  journal  = {{Macromolecular Theory And Simulations}},
  title    = {{Spinodal Decomposition In Ternary Systems With Significantly Different Component Diffusivities}},
  year     = {2007},
  number   = {6},
  pages    = {627-631},
  volume   = {16},
  doi      = {10.1002/mats.200700025},
}

@Article{CHEN1993683,
  author  = {Long-Qing Chen},
  journal = {{Scripta Metallurgica Et Materialia}},
  title   = {{A Computer Simulation Technique For Spinodal Decomposition And Ordering In Ternary Systems}},
  year    = {1993},
  issn    = {0956-716X},
  number  = {5},
  pages   = {683-688},
  volume  = {29},
  doi     = {10.1016/0956-716X(93)90419-S},
  url     = {https://www.sciencedirect.com/science/article/pii/0956716X9390419S},
}

@Article{CHEN19943503,
  author   = {Long-Qing Chen},
  journal  = {{Acta Metallurgica Et Materialia}},
  title    = {{Computer Simulation Of Spinodal Decomposition In Ternary Systems}},
  year     = {1994},
  issn     = {0956-7151},
  number   = {10},
  pages    = {3503-3513},
  volume   = {42},
  doi      = {10.1016/0956-7151(94)90482-0},
  url      = {https://www.sciencedirect.com/science/article/pii/0956715194904820},
}

@Article{eyrie,
  author   = {Eyre, David J.},
  journal  = {{SIAM Journal On Applied Mathematics}},
  title    = {{Systems Of Cahn-Hilliard Equations}},
  year     = {1993},
  number   = {6},
  pages    = {1686-1712},
  volume   = {53},
  doi      = {10.1137/0153078},
}

@Article{owaisprm,
  author    = {Ahmad, Owais and Kumar, Naveen and Mukherjee, Rajdip and Bhowmick, Somnath},
  journal   = {{Phys. Rev. Mater.}},
  title     = {{Accelerating Microstructure Modeling Via Machine Learning: A Method Combining Autoencoder And ConvLSTM}},
  year      = {2023},
  pages     = {083802},
  volume    = {7},
  doi       = {10.1103/PhysRevMaterials.7.083802},
  issue     = {8},
  numpages  = {9},
  publisher = {American Physical Society},
}

@Article{jwcahn_jehilliard,
  author  = {Cahn, John W and Hilliard, John E},
  journal = {{The Journal Of Chemical Physics}},
  title   = {{{Free Energy Of A Nonuniform System. I. Interfacial Free Energy}}},
  year    = {1958},
  number  = {2},
  pages   = {258--267},
  volume  = {28},
  doi     = {10.1063/1.1744102},
}

@Article{bhattacharyya2003study,
  author    = {Bhattacharyya, Saswata and Abinandanan, TA},
  journal   = {{Bulletin Of Materials Science}},
  title     = {{A Study Of Phase Separation In Ternary Alloys}},
  year      = {2003},
  pages     = {193--197},
  volume    = {26},
  doi       = {10.1007/bf02712812},
  publisher = {Springer},
}

@Article{kramer1984interdiffusion,
  author    = {Kramer, Edward J and Green, Peter and Palmstr{\o}m, Christopher J},
  journal   = {{Polymer}},
  title     = {{Interdiffusion And Marker Movements In Concentrated Polymer-polymer Diffusion Couples}},
  year      = {1984},
  number    = {4},
  pages     = {473--480},
  volume    = {25},
  doi       = {10.1016/0032-3861(84)90205-2},
  publisher = {Elsevier},
}

@Article{ghosh2017particles,
  author    = {Ghosh, Supriyo and Mukherjee, Arnab and Abinandanan, TA and Bose, Suryasarathi},
  journal   = {{Physical Chemistry Chemical Physics}},
  title     = {{Particles With Selective Wetting Affect Spinodal Decomposition Microstructures}},
  year      = {2017},
  number    = {23},
  pages     = {15424--15432},
  volume    = {19},
  doi       = {10.1039/c7cp01816a},
  publisher = {Royal Society of Chemistry},
}

@article{lee2012practically,
  title = {{A Practically Unconditionally Gradient Stable Scheme For The N-component Cahn--Hilliard System}},
  author={Lee, Hyun Geun and Choi, Jeong-Whan and Kim, Junseok},
  journal = {{Physica A: Statistical Mechanics And Its Applications}},
  volume={391},
  number={4},
  pages={1009--1019},
  year={2012},
  publisher={Elsevier}
}

@Article{cahn1961spinodal,
  author    = {Cahn, John W},
  journal   = {{Acta Metallurgica}},
  title     = {{On Spinodal Decomposition}},
  year      = {1961},
  number    = {9},
  pages     = {795--801},
  volume    = {9},
  doi       = {10.1016/0001-6160(61)90182-1},
  publisher = {Elsevier},
}

@Article{Bostanabad2018,
  author          = {Bostanabad, Ramin and Zhang, Yichi and Li, Xiaolin and Kearney, Tucker and Brinson, L. Catherine and Apley, Daniel W. and Liu, Wing Kam and Chen, Wei},
  journal         = {{Progress In Materials Science}},
  title           = {{{Computational Microstructure Characterization And Reconstruction: Review Of The State-of-the-art Techniques}}},
  year            = {2018},
  issn            = {0079-6425},
  pages           = {1--41},
  volume          = {95},
  doi             = {10.1016/J.PMATSCI.2018.01.005},
  publisher       = {Pergamon},
}

@Article{Muranushi_2012,
  author   = {Muranushi, Takayuki},
  journal  = {{Computational Science \& Discovery}},
  title    = {{{Paraiso: An Automated Tuning Framework For Explicit Solvers Of Partial Differential Equations}}},
  year     = {2012},
  number   = {1},
  pages    = {15003},
  volume   = {5},
  doi      = {10.1088/1749-4699/5/1/015003},
}

@Article{vondrous2014parallel,
  author    = {Vondrous, Alexander and Selzer, Michael and H{\"o}tzer, Johannes and Nestler, Britta},
  journal   = {{The International Journal Of High Performance Computing Applications}},
  title     = {{Parallel Computing For Phase-field Models}},
  year      = {2014},
  number    = {1},
  pages     = {61--72},
  volume    = {28},
  doi       = {10.1177/1094342013490972},
  publisher = {Sage Publications Sage UK: London, England},
}

@Article{miyoshi2017ultra,
  author    = {Miyoshi, Eisuke and Takaki, Tomohiro and Ohno, Munekazu and Shibuta, Yasushi and Sakane, Shinji and Shimokawabe, Takashi and Aoki, Takayuki},
  journal   = {{NPJ Computational Materials}},
  title     = {{Ultra-large-scale Phase-field Simulation Study Of Ideal Grain Growth}},
  year      = {2017},
  number    = {1},
  pages     = {25},
  volume    = {3},
  doi       = {10.1038/s41524-017-0029-8},
  publisher = {Nature Publishing Group UK London},
}

@Article{montes2021accelerating,
  author    = {Montes de Oca Zapiain, David and Stewart, James A and Dingreville, R{\'e}mi},
  journal   = {{NPJ Computational Materials}},
  title     = {{Accelerating Phase-field-based Microstructure Evolution Predictions Via Surrogate Models Trained By Machine Learning Methods}},
  year      = {2021},
  number    = {1},
  pages     = {1--11},
  volume    = {7},
  doi       = {10.1038/s41524-020-00471-8},
  publisher = {Nature Publishing Group},
}

@Article{hu2022accelerating,
  author    = {Hu, C and Martin, S and Dingreville, R},
  journal   = {{Computer Methods In Applied Mechanics And Engineering}},
  title     = {{Accelerating Phase-field Predictions Via Recurrent Neural Networks Learning The Microstructure Evolution In Latent Space}},
  year      = {2022},
  pages     = {115128},
  volume    = {397},
  doi       = {10.1016/j.cma.2022.115128},
  publisher = {Elsevier},
}

@Article{findik1993sidebands,
  author    = {Findik, F},
  journal   = {{Journal Of Materials Science Letters}},
  title     = {{Sidebands In Spinodal Cu-Ni-Cr Alloys And Lattice Parameters Inquiries}},
  year      = {1993},
  number    = {5},
  pages     = {338--342},
  volume    = {12},
  doi       = {10.1007/bf01910098},
  publisher = {Springer},
}

@Article{morral1971spinodal,
  author    = {Morral, JE and Cahn, JW},
  journal   = {{Acta Metallurgica}},
  title     = {{Spinodal Decomposition In Ternary Systems}},
  year      = {1971},
  number    = {10},
  pages     = {1037--1045},
  volume    = {19},
  doi       = {10.1016/0001-6160(71)90036-8},
  publisher = {Elsevier},
}

@Article{de1972analysis,
  author    = {De Fontaine, D},
  journal   = {{Journal Of Physics And Chemistry Of Solids}},
  title     = {{An Analysis Of Clustering And Ordering In Multicomponent Solid Solutions-I. Stability Criteria}},
  year      = {1972},
  number    = {2},
  pages     = {297--310},
  volume    = {33},
  doi       = {10.1016/0022-3697(72)90011-x},
  publisher = {Elsevier},
}

@Article{de1973analysis,
  author    = {De Fontaine, D},
  journal   = {{Journal Of Physics And Chemistry Of Solids}},
  title     = {{An Analysis Of Clustering And Ordering In Multicomponent Solid Solutions-II Fluctuations And Kinetics}},
  year      = {1973},
  number    = {8},
  pages     = {1285--1304},
  volume    = {34},
  doi       = {10.1016/0038-1098(73)90679-0},
  publisher = {Elsevier},
}

@Article{de1979configurational,
  author    = {De Fontaine, Didier},
  journal   = {{Solid State Physics}},
  title     = {{Configurational Thermodynamics Of Solid Solutions}},
  year      = {1979},
  pages     = {73--274},
  volume    = {34},
  doi       = {10.1016/s0081-1947(08)60360-4},
  publisher = {Elsevier},
}

@Article{rios2011spinodal,
  author    = {Rios, Orlando and Ebrahimi, Fereshteh},
  journal   = {{Intermetallics}},
  title     = {{Spinodal Decomposition Of The $\gamma$-phase Upon Quenching In The Ti--Al--Nb Ternary Alloy System}},
  year      = {2011},
  number    = {1},
  pages     = {93--98},
  volume    = {19},
  doi       = {10.1016/j.intermet.2010.09.014},
  publisher = {Elsevier},
}

@article{Zhang2018,
  author  = {Zhang, Chi and Xie, Zhuohong and He, Xin and Liang, Ping and Zeng, Qingguang and Zhang, Zhonghua},
  title   = {Fabrication and Characterization of Nanoporous Cu--Sn Intermetallics via Dealloying of Ternary Mg--Cu--Sn Alloys},
  journal = {CrystEngComm},
  year    = {2018},
  volume  = {20},
  pages   = {6900--6908},
  doi     = {10.1039/C8CE01328D}
}

@article{Song2015,
  author  = {Song, Tingting and Yan, Ming and Shi, Zhiming and Atrens, Andre and Qian, Ma},
  title   = {Electrochemical Dealloying of a Ternary Al67Cu18Sn15 Alloy Compared with that of a Binary Al75Cu25 Alloy},
  journal = {ECS Transactions},
  year    = {2015},
  volume  = {66},
  number  = {40},
  pages   = {23--30},
  doi     = {10.1149/06640.0023ecst}
}

@article{Geslin2015,
  author  = {Geslin, Pierre-Antoine and McCue, Ian and Gaskey, Bernard and Erlebacher, Jonah and Karma, Alain},
  title   = {Topology-Generating Interfacial Pattern Formation during Liquid Metal Dealloying},
  journal = {Nature Communications},
  year    = {2015},
  volume  = {6},
  pages   = {8887},
  doi     = {10.1038/ncomms9887}
}

@article{Qi2013,
  author  = {Qi, Zhen and Weissm{\"u}ller, J{\"o}rg},
  title   = {Hierarchical Nested-Network Nanostructure by Dealloying},
  journal = {ACS Nano},
  year    = {2013},
  volume  = {7},
  number  = {7},
  pages   = {5948--5954},
  doi     = {10.1021/nn4021345}
}

@article{Kim2009,
  author  = {Kim, Yun Ho and Yoon, Dong Ki and Jung, Hee-Tae},
  title   = {Recent Advances in the Fabrication of Nanotemplates from Supramolecular Self-Organization},
  journal = {Journal of Materials Chemistry},
  year    = {2009},
  volume  = {19},
  pages   = {9091--9102},
  doi     = {10.1039/B910496H}
}

\end{document}